%% file: main-icml.tex
\documentclass{article}

\usepackage{microtype}
\usepackage{graphicx}
\usepackage{subcaption}
\usepackage{booktabs} 
\usepackage{hyperref}
\usepackage{enumitem}

\PassOptionsToPackage{sort}{natbib}

\usepackage[preprint]{icml2026}

\usepackage{subcaption}
\usepackage{graphicx}
\usepackage{amsmath}
\usepackage{amssymb}
\usepackage{mathtools}
\usepackage{amsthm}
\usepackage{algorithm}
\usepackage{algorithmic}
\usepackage[capitalize,noabbrev]{cleveref}
\usepackage{multirow} 
\usepackage[most]{tcolorbox}
\usepackage{float}
\usepackage{tabularx, makecell}
\usepackage{pifont}
\newtcolorbox{promptbox}[1]{
    enhanced,
    colback=blue!5,
    colframe=blue!45!black,
    colbacktitle=blue!45!black,
    coltitle=white, 
    fonttitle=\bfseries\sffamily,
    attach boxed title to top left={yshift=-3mm, xshift=3mm},
    boxed title style={
        sharp corners, 
        rounded corners=southeast, 
        arc=2mm, 
        titlerule=0pt
    },
    rounded corners,
    arc=2mm,
    boxrule=1pt,  
    left=10pt, right=10pt, top=12pt, bottom=10pt, 
    fontupper=\rmfamily,
    title=#1,
    breakable
}
\newcolumntype{C}[1]{>{\centering\arraybackslash}p{#1}}

\usepackage{soul}
\usepackage{xcolor}

\definecolor{boxTitleBg}{RGB}{230, 230, 230}
\definecolor{boxFrame}{RGB}{100, 100, 100}
\definecolor{bgQuery}{RGB}{235, 245, 250}
\definecolor{bgChunkA}{RGB}{240, 255, 240}
\definecolor{bgChunkB}{RGB}{255, 245, 235}
\definecolor{bgResult}{RGB}{250, 250, 250}
\definecolor{bgAnswer}{RGB}{255, 250, 240}

\definecolor{colorA}{RGB}{178, 223, 219}      
\definecolor{colorB}{RGB}{255, 205, 210}      
\definecolor{colorBridge}{RGB}{187, 222, 251}

\DeclareRobustCommand{\hlA}[1]{{\sethlcolor{colorA}\hl{\textbf{#1}}}}
\DeclareRobustCommand{\hlB}[1]{{\sethlcolor{colorB}\hl{\textbf{#1}}}}
\DeclareRobustCommand{\hlBridge}[1]{{\sethlcolor{colorBridge}\hl{\textbf{#1}}}}

\theoremstyle{plain}

\theoremstyle{definition}

\theoremstyle{remark}

\usepackage[textsize=tiny]{todonotes}
\usepackage{tikz}
\usepackage{amsmath}
\usetikzlibrary{positioning, shapes.geometric, arrows.meta, decorations.pathreplacing, calc}

\newcommand{\eg}{\textit{e.g.}}
\newcommand{\ie}{\textit{i.e.}}
\newcommand{\wrt}{\textit{w.r.t.}}
\newcommand{\etc}{\textit{etc.}}

\newcommand{\ourname}{MergeRAG}

\definecolor{gjr}{HTML}{FF0000}

\newcommand{\thetitle}{Rethinking Retrieval-Augmentation as Synthesis: A Query-Aware Context Merging Approach}
\newcommand{\RETURN}{\STATE \textbf{return} }

\icmltitlerunning{\thetitle}

\begin{document}

\twocolumn[
  \icmltitle{\thetitle}
  \icmlsetsymbol{equal}{*}
  \begin{icmlauthorlist}
    \icmlauthor{Jiarui Guo}{equal,pku,alibaba}
    \icmlauthor{Yuemeng Xu}{equal,pku}
    \icmlauthor{Zongwei Lv}{equal,pku,alibaba}
    \icmlauthor{Yangyujia Wang}{pku} \\
    \icmlauthor{Xiaolin Wang}{pku}
    \icmlauthor{Kan Liu}{alibaba}
    \icmlauthor{Tao Lan}{alibaba}
    \icmlauthor{Lin Qu}{alibaba}
    \icmlauthor{Tong Yang}{pku}
  \end{icmlauthorlist}
  \icmlaffiliation{pku}{Peking University}
  % \icmlaffiliation{pku-math}{School of Mathematical Sciences, Peking University}
  \icmlaffiliation{alibaba}{Alibaba Group}
  \icmlcorrespondingauthor{Kan Liu}{liukan.lk@alibaba-inc.com}
  \icmlcorrespondingauthor{Tong Yang}{yangtong@pku.edu.cn}
  \vskip 0.3in
]

\printAffiliationsAndNotice{\textsuperscript{*}Co-first authors with equal contribution. Listing order is random.}

\input{content/0_abstract_v2}

\input{content/1_introduction}

\input{content/2_related_work}
\input{content/3_algorithm_v2}

\input{content/4_experiments}

\input{content/5_conclusion}

% \section*{Accessibility}

% \section*{Software and Data}

% \section*{Acknowledgements}

% \newpage
% \clearpage
\section*{Impact Statement} 
This paper presents work whose goal is to advance the field of Retrieval-Augmented Generation (RAG). 
There are many potential societal consequences of our work, none which we feel must be specifically highlighted here.

\bibliography{reference}
\bibliographystyle{icml2026}

\newpage
\appendix
\onecolumn
\input{content/6_appendix}

\end{document}

%% file: content/0_abstract_v2.tex
\begin{abstract}
    Retrieval-Augmented Generation (RAG) enables Large Language Models (LLMs) to extend their existing knowledge by dynamically incorporating external information.
    However, practical deployment is fundamentally constrained by the LLM's finite context window, forcing a trade-off between information sufficiency and token consumption. 
    Standard pipelines address this via a \textit{retrieve-then-select} strategy, typically retaining only the top-$k$ chunks based on relevance.
    Nevertheless, this approach is suboptimal: it inherently truncates critical bridging evidence located in the long tail of the relevance distribution, while simultaneously wasting the token budget on semantically redundant high-ranking chunks. 
    
    In this paper, we rethink retrieval-augmentation as a dynamic optimization problem aimed at maximizing information density. 
    We propose \textbf{\ourname{}}, a novel framework that shifts the paradigm from static filtering to query-aware synthesis. 
    \ourname{} employs a scoring agent to restructure retrieved contexts through a dual-pathway mechanism: 
    \textbf{1) Symmetric Merging}, which consolidates weak signals to recover lost bridging evidence; 
    \textbf{2) Asymmetric Merging}, which utilizes entropy-guided anchoring to eliminate redundancy without sacrificing semantic integrity. 
    We further introduce a \textbf{Hierarchical Parallel Merging} strategy that mitigates information loss while maximizing computational parallelism. 
    Extensive experiments on standard benchmarks demonstrate that \ourname{} significantly outperforms state-of-the-art RAG baselines, achieving up to 13.7 points improvement in F1 score and 11.5 points in Exact Match (EM), respectively. 
\end{abstract}

%% file: content/1_introduction.tex
\section{Introduction}

Large Language Models (LLMs) \cite{liu2024deepseek, touvron2023llama, achiam2023gpt} have fundamentally transformed the landscape of artificial intelligence, demonstrating unprecedented capabilities in language understanding, generation and context reasoning. 
However, despite expensive pre-training on vast corpora, LLMs operating in a closed-book setting are susceptible to knowledge obsolescence and hallucinations \cite{mousavi2025llms, ji2023towards}, as they rely solely on static internal parameters. 
To mitigate these limitations, Retrieval-Augmented Generation (RAG) has been proposed to complement the model's frozen weights with external memory \cite{gao2023retrieval, guu2020retrieval, lewis2020retrieval}. 
By augmenting the input with relevant retrieved contexts, RAG allows LLMs to access up-to-date domain knowledge without retraining. 
Consequently, RAG has emerged as the prevalent paradigm for enhancing the factual integrity of LLMs in open-domain scenarios \cite{amirshahi2025evaluating, siriwardhana2023improving}.

\begin{figure*}
    \centering
    \includegraphics[width=.95\linewidth]{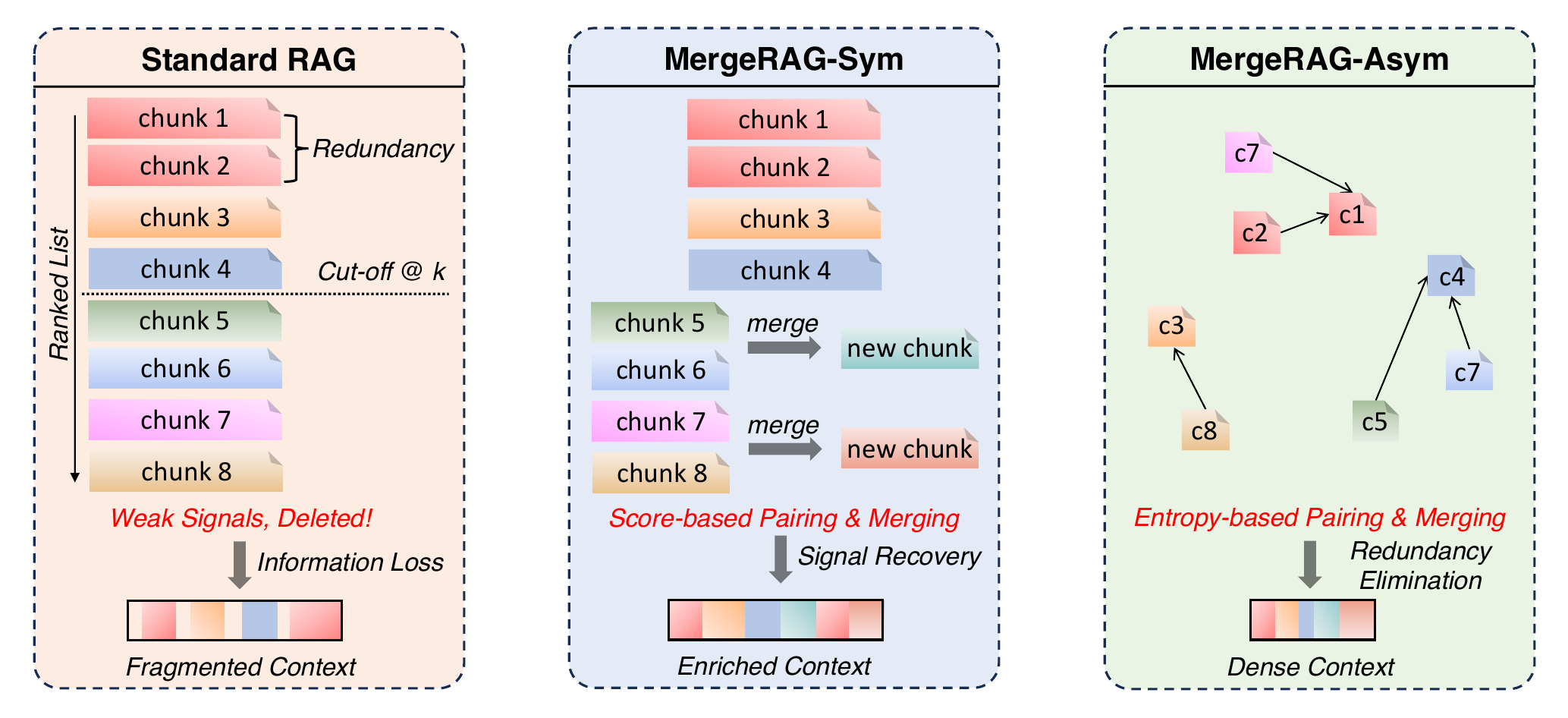}
    \caption{Comparison between standard RAG solutions and \ourname{}.}
    \label{fig:example}
\end{figure*}

Nevertheless, the practical application of RAG is constrained by the inherent limitations of the LLM's context window \cite{gao2025u, li2025long}. 
To operate within these constraints, the standard practice adopts a \textit{retrieve-then-select} paradigm, typically retaining only the top-$k$ chunks based on semantic similarity scores \cite{karpukhin2020dense, izacard2023atlas}. 
While efficient in many cases, this approach inherently suffers from information loss by truncating bridging evidence in low-scoring chunks, and results in a fragmented context by ignoring inter-chunk relationships. 
Furthermore, this greedy selection often introduces significant semantic redundancy, as multiple high-ranking chunks may convey repetitive information. 
This redundancy wastefully consumes the limited token budget, displacing unique details that are essential for comprehensive reasoning.
Alternatively, some approaches attempt to improve context organization via offline pre-processing (\eg{} static clustering or summarization) \cite{tao2025treerag, fatehkia2024t, sarthi2024raptor}. 
However, since these structures are built prior to receiving the user's input, these query-agnostic methods lack the flexibility to align with the specific semantic nuances of the dynamic query, leading to suboptimal relevance.

To bridge this gap, we propose a paradigm shift from simply \textit{filtering} existing chunks to dynamically \textit{synthesizing} them in a query-aware manner.
Our core insight is that the optimal information organization is not static but intrinsically tied to the specific user intent: 
the ideal context for a complex query is rarely a single perfect document, but rather a composite of multiple, often fragmentary clues, that must be pieced together at runtime \cite{xiong2020answering, yang2018hotpotqa}. 
Therefore, instead of discarding lower-rank chunks, the system should allow these fragments to \textit{merge} into coherent units based on their actual relevance to the query. 
By iteratively fusing related segments, we can construct a context that maximizes information density while maintaining narrative coherence. 
This approach ensures that the LLM receives a structured reasoning path tailored to the specific query, rather than a disjointed bag of retrieval results, effectively overcoming the limitations of static selection \cite{wei2022chain, zhang2022automatic}.

To materialize this insight, we propose \textbf{\ourname{}}, a novel framework to achieve the optimal trade-off between maximizing context relevance and minimizing context length.
As shown in Figure \ref{fig:example}, \ourname{} employs a scoring agent to dynamically restructure the retrieved chunks via two complementary strategies: 
\textbf{1) Symmetric Merging}, which targets the long-tail of the retrieved chunks by consolidating weak signals to recover critical evidence rather than directly discarding them;
\textbf{2) Asymmetric Merging}, which maximizes information density via entropy-guided anchoring, fusing predictable fragments onto host chunks to eliminate redundancy. 
We further introduce a \textbf{Hierarchical Parallel Merging} algorithm that restructures context construction into a logarithmic-depth tree, minimizing recursive generation errors while maximizing inference parallelism.
Our experiments demonstrate that \ourname{} achieves a retrieval improvement of up to 1.51$\times$ on multi-hop reasoning benchmarks compared to existing RAG solutions. 
% \footnote{
% All source code of \ourname{} is available on GitHub anonymously via \url{https://github.com/MergeRAG/MergeRAG-src}. 
% }

In general, the contributions of this paper are summarized as follows: 
\begin{itemize}[leftmargin=1em]
    \item{\textbf{Theory:}} 
    We conceptualize RAG context construction as a dual-objective optimization problem. 
    We argue that an effective system must jointly maximize semantic relevance and minimize token usage, rather than treating them as independent constraints or relying on static selection.
    \item{\textbf{Algorithm:}} 
    To address this, we propose the \ourname{} framework. 
    This query-aware algorithm shifts the paradigm from filtering to dynamic synthesis, and implements Symmetric Merging for signal boosting and Asymmetric Merging for redundancy reduction. 
    \item{\textbf{Experiment:}}
    We conduct extensive experiments across diverse benchmarks. 
    The results demonstrate that \ourname{} significantly outperforms state-of-the-art baselines, achieving substantial improvements in both retrieval fidelity and generation quality.
\end{itemize}

%% file: content/2_related_work.tex
\section{Related Work}

The advancement of retrieval-augmented generation (RAG) has fundamentally shifted the landscape of LLMs from static parametric knowledge bases to dynamic, open-book reasoning engines. 
Since the efficacy of a RAG system depends critically on the quality, density, and structural organization of the retrieved context, we review the literature regarding these three dimensions.

\subsection{Context Selection}

Context selection is the foundational layer of RAG, which aims to filter out chunks irrelevant to the user query and retain only the most significant ones \cite{lewis2020retrieval}. 
Traditionally, RAG pipelines quantify relevance using bi-encoders or sparse retrievers \cite{reimers2019sentence, robertson2009probabilistic, karpukhin2020dense}. 
Then, a fixed top-$k$ truncation is applied to select the highest-scoring chunks as the input context. 
More recently, LLMs have been deployed as effective reranking agents, leveraging their superior reasoning capability to refine chunk ordering and outperform supervised baselines \cite{sun2023chatgpt, zhang2023rankinggpt, ma2024fine}. 

To enhance flexibility, recent research has shifted from fixed top-$k$ truncation to adaptive mechanisms that adjust the retrieval scope according to query complexity. 
For example, Dynamic Passage Selection (DPS) \cite{meng2025ranking} determines the optimal set of passages in a query-dependent manner, rather than adhering to a pre-defined $k$.
Furthermore, active strategies have also emerged: FLARE \cite{jiang2023active} iteratively decides when to retrieve based on generation confidence, while Self-RAG \cite{asai2024self} trains a model to generate reflection tokens to autonomously critique whether more information is needed. 

Despite the advancements, these approaches primarily position context construction as a selection problem. 
They focus on \textit{which} or \textit{how many} chunks to retain, but treat the retrieved segments as atomic, immutable units. 
Therefore, they do not alter the internal structure of the chunks or synthesize information across them. 
This leads to a dual failure: significant redundancy remains within the selected subset, while critical bridging evidence located in the long tail is inevitably discarded due to rigid truncation.

\subsection{Context Compression}

To mitigate window constraints, research has pivoted toward compressing retrieved information. 
Token-level approaches like Selective Context \cite{li2023compressing} and LongLLMLingua \cite{jiang2024longllmlingua} utilize perplexity-based metrics to prune tokens with low semantic value, and RECOMP \cite{xu2023recomp} trains neural models to generate abstractive summaries. 
Recent works like xRAG \cite{cheng2024xrag} and COCOM \cite{rau2025context} project documents into dense vectors to bypass the verbosity of natural language. 
In addition, SARA \cite{jin2025sara} adopts a hybrid strategy to selectively preserve critical evidence in natural language while compressing supplementary context into vectors. 

Despite the efficiency of compression, these solutions primarily position context optimization as a subtractive task. 
They focus on how to condense textual elements to meet token constraints, yet inherently function as lossy operators. 
Consequently, by aggressively pruning tokens or abstracting content, they risk discarding critical fine-grained entities, sacrificing semantic fidelity for the sake of brevity.

\subsection{Knowledge Aggregation}

Beyond linear selection and compression, a distinct line of research focuses on pre-organizing retrieved information to capture inter-chunk relationships through offline pre-processing. 
Approaches like RAPTOR \cite{sarthi2024raptor} employ recursive clustering to construct hierarchical trees, allowing for retrieval at different levels of abstraction. 
Similarly, TreeRAG \cite{tao2025treerag, fatehkia2024t, li2025cft} introduces a tree-chunking mechanism to map long documents into hierarchical structures to preserve semantic integrity and inter-chunk dependencies. 
GraphRAG \cite{edge2024local} converts corpora into knowledge graphs to model entity dependencies and generate community summaries. 

However, these structural aggregations are generally static and query-agnostic, as these structures are solidified during the offline indexing phase and do not adapt to the specific user query. 
As a result, this rigid pre-organization lacks the flexibility to dynamically synthesize evidence for ad-hoc lines of reasoning, often retrieving entire pre-built branches that may contain irrelevant noise relative to the user's immediate intent.

%% file: content/3_algorithm_v2.tex
\section{Methods}

\subsection{Overview and Problem Formulation}

The core challenge in Retrieval-Augmented Generation (RAG) lies in resolving the intrinsic tension between informational sufficiency and representation complexity. 
While expanding the context window increases the likelihood of encompassing the correct answer (sufficiency), it simultaneously incurs significant computational overhead and introduces semantic noise that dilutes the model's attention (complexity). 
We frame this tension as a rate-distortion optimization problem, rigorously grounded in the \textbf{Information Bottleneck (IB)} principle \cite{tishby2000information}.
% We formulate the context construction task as an optimization problem grounded in the \textbf{Information Bottleneck (IB) principle} \cite{tishby2000information}.

Formally, consider a RAG scenario where a user query $q$ is answered utilizing a set of retrieved document chunks. 
Let $\mathcal{C} = \{c_1, c_2, \dots, c_N\}$ denote the set of retrieved chunks. 
Each chunk $c_i$ is associated with a query-relevance score $s_i = S(c_i; q)$ and a token length $l_i = l(c_i)$.
Our objective is to construct a compressed context $\mathcal{S}$ that maximizes the task-relevant information regarding $q$ while minimizing its token length. 
This is formalized as the Lagrangian relaxation:
\begin{equation}
\label{eq:opt}
    \max_{\mathcal{S}} \mathcal{L} = \underbrace{I(\mathcal{S}; q)}_{\text{Relevance (Accuracy)}} - \beta \cdot \underbrace{\sum_{c\in \mathcal{S}}l(c)}_{\text{Compression (Cost)}},
\end{equation}
where $I(\cdot;\cdot)$ denotes mutual information, and $\beta>0$ is a Lagrange multiplier controlling the trade-off.
However, directly optimizing Equation \ref{eq:opt} is computationally intractable due to the combinatorial nature of the discrete text space and the lack of explicit probability distributions for $I(\mathcal{S}; q)$.

\textbf{Standard RAG Approximation:} Traditional RAG solutions employ a static selection strategy, identifying the top-$k$ chunks S based on their relevance scores:
\[
\mathcal{S} = \underset{c \in \mathcal{C}}{\text{top-}k} \, S(c;q). 
\] 
Since existing pipelines typically pre-segment documents into chunks of relatively fixed size, this greedy selection process is mathematically equivalent to maximizing total relevance under a context window constraint $L$: 
\[ \mathcal{S} = \underset{\mathcal{S}' \subseteq \mathcal{C}}{\arg\max} \sum_{c \in \mathcal{S}'} S(c;q) \quad \text{s.t.} \quad \sum_{c \in \mathcal{S}'} l(c) \leq L. 
\]
However, while computationally efficient, this truncation leads to significant information loss and context fragmentation, as it strictly discards lower-ranked chunks that may contain essential bridging information.

\textbf{Our Dynamic Approximation:}
To better approximate the objective, \ourname{} introduces a dynamic merging framework that addresses Equation \ref{eq:opt} through two distinct strategies: 
Symmetric Merging focuses on the first term (relevance) by consolidating weak signals to recover lost information, whereas Asymmetric Merging targets the second term (cost) by leveraging conditional entropy to compress redundant context.

\subsection{Symmetric Merging: Consolidating Weak Signals}

% \paragraph{Semantic Fusion: Query-Conditional Distillation.}
% Qualitatively, we implement the fusion operator $\mathcal{M}$ as a \textbf{query-conditional distillation process}. 
% We leverage the LLM as a \textit{semantic filter} steered by the prompt. 
% This mechanism controls the information flow to explicitly maximize the \textbf{mutual information} $I(q; c^*)$ between the query and the merged content, while discarding the irrelevant noise often found in low-ranking chunks. 
% By forcing the model to reconstruct the joint information within a constrained length, we effectively distill the scattered weak signals into a compact, high-density unit, simultaneously achieving token compression and signal enhancement.

The first objective in Equation \ref{eq:opt} is to maximize the mutual information $I(\mathcal{S};q)$. 
Standard selection assumes that chunks with low scores have negligible mutual information, \ie{}, $I(c;q) \approx 0$ when its score $S(c;q)$ is low. 
However, for complex queries, the answer often relies on the aggregation of weak signals located in the long tail of the relevance distribution. 
To address this, \textbf{Symmetric Merging} strategy targets the weakest signals: 
by consolidating these low-scoring chunks, we aim to recover the latent mutual information that would otherwise be lost via truncation. 

\textbf{Iterative Merging Process:}
We implement this strategy as an iterative reduction process. 
Let $\mathcal{C}^{(t)}$ denote the set of chunks at iteration $t$, and $\mathcal{C}^{(0)}=\mathcal{C}$. 
The process repeats the following steps until the set size reaches a target budget $L$, \ie{}, $\sum_{c\in \mathcal{C}^{(t)}}l(c) \leq L$.

\textbf{1) Selection.} 
In each step, we identify the pair $(c_i^*, c_j^*)$ with minimum cumulative relevance score, \ie{}, 
\[
(c_i^*, c_j^*) \leftarrow \underset{c_i, c_j \in \mathcal{C}^{(t)}, i\neq j}{\arg\min} \left\{S(c_i;q)+ S(c_j;q)\right\}. 
\]
This greedy strategy acts as a bottom-up denoising process, preserving high-confidence anchors while attempting to consolidate the tail of the distribution into detectable signals.

\textbf{2) Fusion.}
The selected pair is processed by a fusion operator $\mathcal{M}$, which is not merely a concatenation function but a semantic compressor. 
Guided by the user query $q$, we instruct the LLM to synthesize the disparate facts from $c_i^*$ and $c_j^*$ into a coherent unit $c^*$, selectively preserving only the information necessary for answering $q$, \ie{},
\[
c^* \leftarrow \mathcal{M}(c_i^*, c_j^*; q).
\]

\textbf{3) Update.}
Once $c^*$ is generated, we remove the original pair from $\mathcal{C}$ and replace them with $c^*$, \ie{},
\[
\mathcal{C}^{(t+1)} \leftarrow \left(\mathcal{C}^{(t)} \setminus \{c_i^*, c_j^*\}\right) \cup \{c^*\}.
\]
We then calculate its length $l(c^*)$ and relevance score $s(c^*;q)$ for later iteration.

\subsection{Asymmetric Merging: Entropy-Guided Anchoring}

\begin{figure}[!t]
	\centering
	\begin{subfigure}[t]{0.25\textwidth}{
			\includegraphics[width=\textwidth]{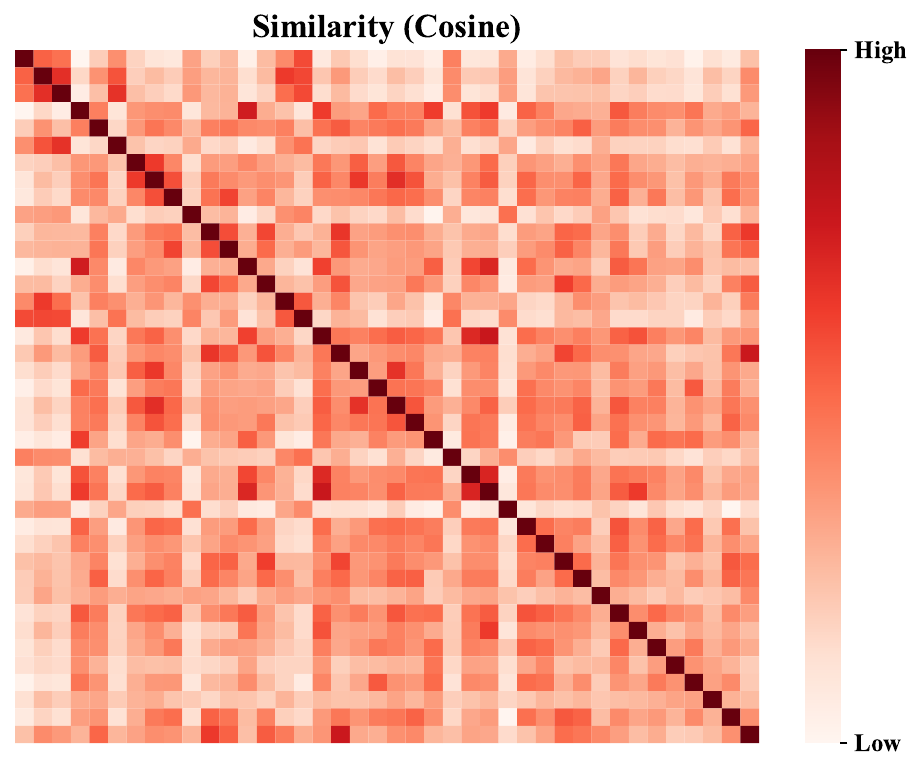}
			\label{fig:cos}
            \caption{Cosine similarity}
		}
	\end{subfigure}%
	\begin{subfigure}[t]{0.25\textwidth}{
			\includegraphics[width=\textwidth]{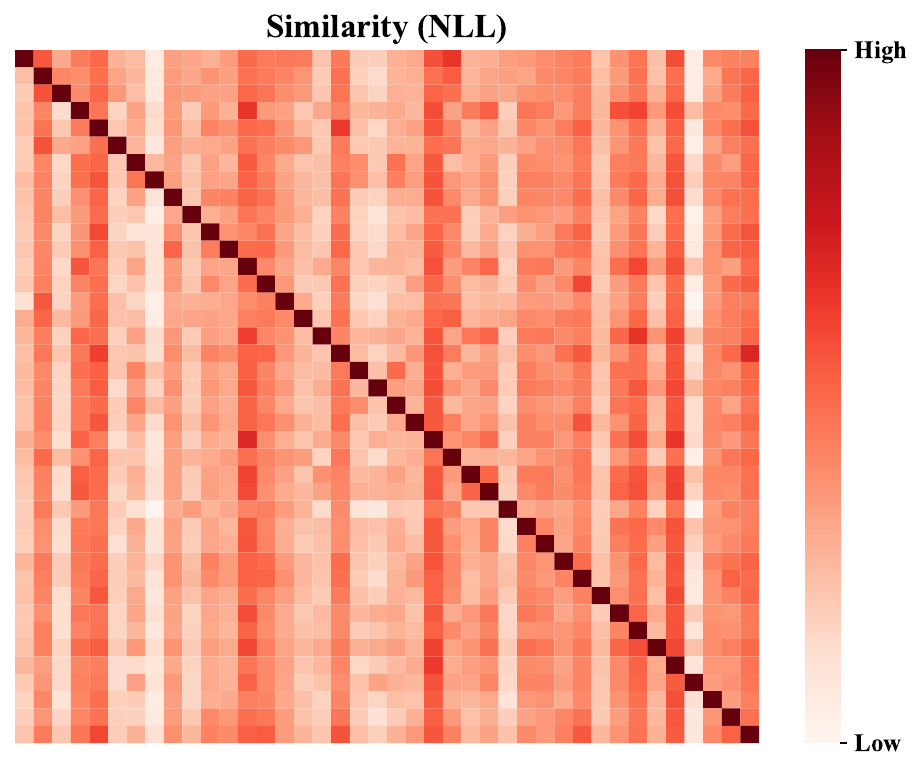}
			\label{fig:nll}
            \caption{NLL similarity}
		}
	\end{subfigure}
	\caption{Similarity examples of chunks. The chunks are sorted by relevance score, with chunks to the top-left corner having high relevance scores. The cell at position $(i,j)$ denotes the similarity between chunks $c_i$ and $c_j$.}
    \label{fig:simi}
\end{figure}

While symmetric merging effectively recovers lost signals, high-recall retrieval inevitably introduces semantic redundancy, where multiple chunks convey overlapping information.
To investigate the distribution of this redundancy, we visualize the pairwise chunk correlations of a representative example from the MuSiQue dataset \cite{trivedi2022musique}. 
As shown in Figure \ref{fig:simi}, high-similarity clusters are often scattered rather than concentrated at the bottom. 
Consequently, simply merging the two lowest-scoring chunks is sometimes suboptimal, as a low-scoring chunk is often redundant to a high-scoring `anchor' rather than its low-scoring neighbor. 

\begin{figure}[!t]
    \centering
    \includegraphics[width=\linewidth]{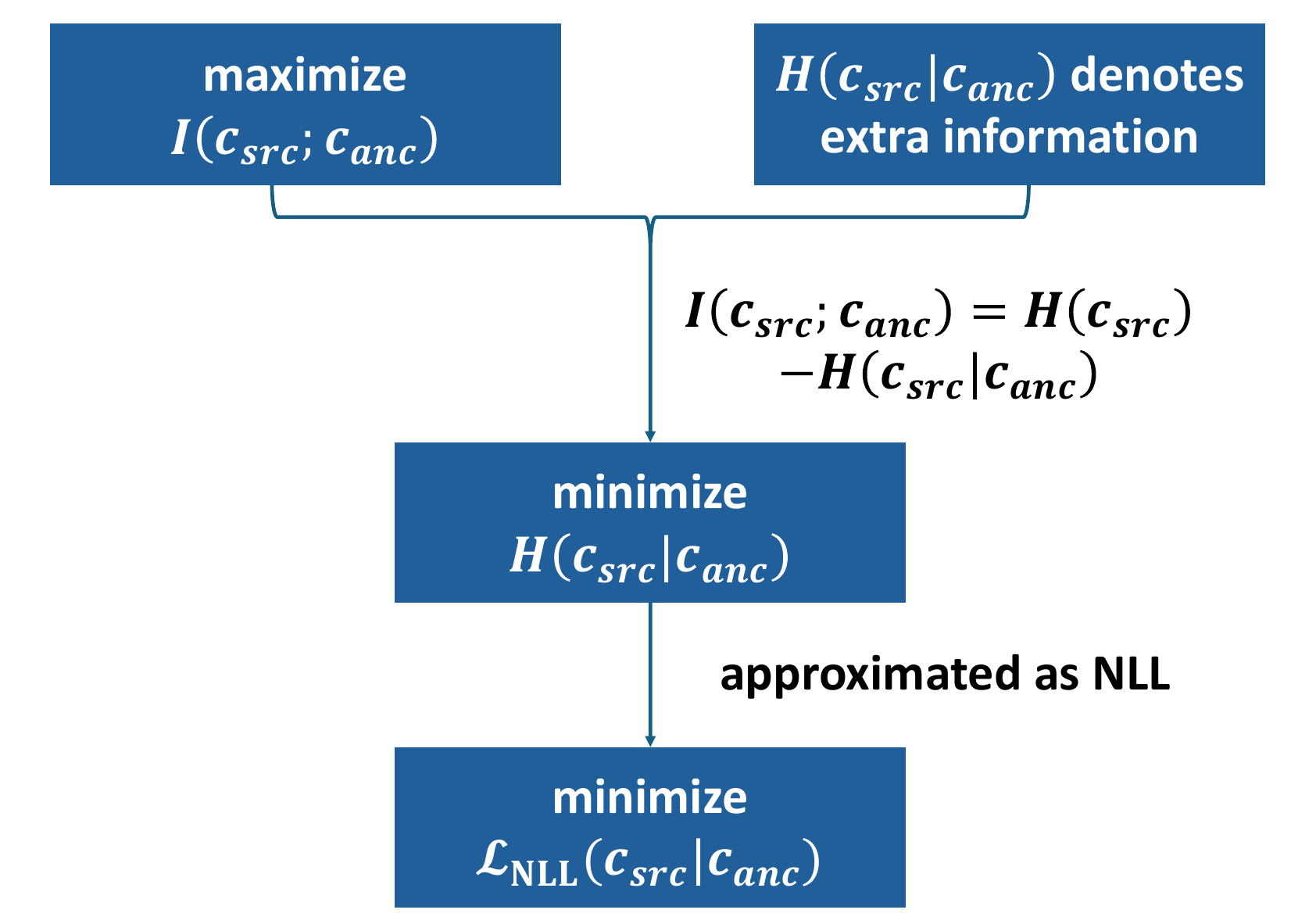}
    \caption{The information-theoretic formulation of Asymmetric Merging.
    We formulate redundancy elimination as maximizing the mutual information $I(c_{src};c_{anc})$. 
    Leveraging the entropy identity, it is equivalent to minimizing the conditional entropy $H(c_{src}|c_{anc})$, which represents the extra description length of $c_{src}$ given $c_{anc}$.
    The conditional entropy is further approximated by minimizing the Negative Log-Likelihood $\mathcal{L}_{\text{NLL}}(c_{src}|c_{anc})$. 
    }
    \label{fig:asym}
\end{figure}

To address this, we aim to minimize token length by eliminating such repetition (Equation \ref{eq:opt}).
We formalize this by maximizing the mutual information $I(c_{src}; c_{anc})$ to capture the semantic overlap between a source chunk $c_{src}$ and an anchor chunk $c_{anc}$.
Leveraging the identity $I(c_{src}; c_{anc}) = H(c_{src}) - H(c_{src}|c_{anc})$, maximizing mutual information for a fixed source is equivalent to minimizing the conditional entropy $H(c_{src}|c_{anc})$.
Crucially, Shannon's Source Coding Theorem \cite{shannon1948mathematical} also establishes this conditional entropy as the lower bound for the code length required to represent $c_{src}$ given $c_{anc}$.
Translating this to the language modeling context, we introduce \textbf{Asymmetric Merging}, which utilizes the conditional Negative Log-Likelihood (NLL) as an empirical estimator of the conditional entropy.
By minimizing the NLL, we mathematically minimize the lower bound of the effective description length, thereby identifying the semantic anchor that most efficiently subsumes the source chunk for lossless compression. 
The theoretical motivation of Asymmetric Merging is shown in Figure \ref{fig:asym}.

\textbf{Iterative Merging Process:} 
Similar to the symmetric strategy, this process iteratively reduces the context size until it fits the target budget. In each iteration $t$:

\textbf{1) Selection.}
Instead of selecting \textit{two} low-scoring chunks directly in Symmetric Merging, the asymmetric variant selects only \textit{one} chunk with minimum score. 
Subsequently, to identify the optimal host for this fragment, we employ an entropy-based metric by computing the Negative Log-Likelihood (NLL) of the selected chunk $c_i^*=(x_1, x_2, \cdots, x_{l(c_i^*)})$ conditioned on each candidate anchor $c_j$, \ie{}, 
\[
\mathcal{L}_{\text{NLL}}(c_i^*|c_j) = -\frac{1}{l(c_i^*)} \sum_{t=1}^{l(c_i^*)}  \log P_{\theta}(x_t|c_j, x_{<t}),
\]
where $P_\theta$ is the probability distribution of the underlying language model, and $x_{<t}$ denotes the preceding tokens within $c_i^*$. 
A lower $\mathcal{L}_{\text{NLL}}$ value indicates that $c_j$ provides strong contextual support for $c_i^*$, making it an ideal anchor for assimilation. 
Therefore, we select the anchor $c_{j}^*$ that minimizes this value, \ie{}, 
\[
c_{j}^* \leftarrow \underset{c_j \in \mathcal{C}^{(t)}, j\neq i}{\arg\min} \mathcal{L}_{\text{NLL}}(c_{i}^*|c_j), 
\]
and pass $(c_i^*, c_j^*)$ to the LLM for subsequent fusion.

\textbf{2) Fusion:} 
Unlike the egalitarian synthesis employed in Symmetric Merging, the fusion process here is strictly directional.
We designate the identified anchor $c_j^*$ as the primary host, and the low-scoring chunk $c_i^*$ as an auxiliary source. 
The fusion operator $\mathcal{M}$ then performs a query-guided augmentation: 
the LLM is instructed to preserve the structural integrity of the host $c_j^*$, while selectively extracting and injecting query-relevant details from $c_i^*$. 
Effectively, $c_j^*$ assimilates salient signals from $c_i^*$ while filtering out the remaining noise. 
This ensures that the high-quality context of the anchor is enriched rather than diluted in the resulting merged chunk $c^*$. 

\textbf{3) Update:} $c_i^*$ is explicitly removed to achieve compression, while the anchor $c_j^*$ is updated with the fused content:
\[
\mathcal{C}^{(t+1)} \leftarrow (\mathcal{C}^{(t)} \setminus \{c_{i}^*, c_{j}^*\}) \cup \{c^*\}.
\]

\subsection{Hierarchical Parallel Merging}

While the proposed Symmetric and Asymmetric Merging methods define \textit{which} chunks to merge, a naive sequential implementation incurs two major drawbacks.
First, reducing the chunk set from size $N$ to $M$ sequentially necessitates $N-M$ consecutive inference steps, introducing significant latency in real-time scenarios.
Second, sequential merging forces early-merged chunks to undergo repeated rounds of LLM regeneration, where a single source chunk may be recursively processed up to $N-M$ times in the worst-case scenario. 
Since each fusion step introduces a non-zero probability of hallucination or information loss, this deep recursive processing progressively dilutes the original semantic signal, leading to context degradation.

To address these issues, we introduce a \textbf{Hierarchical Parallel Merging} strategy. 
Instead of processing a pair of chunks in each iteration, we view the context construction as a tree-structured reduction process, where disjoint pairs are fused concurrently to progressively condense the information.

\textbf{Rationale:}
In iteration $t$, we treat the current chunk set $\mathcal{C}^{(t)}$ as a pool of candidates. 
We maximize parallelism by identifying and merging \textit{all} valid pairs simultaneously in a single layer, rather than individually merging one pair. 
This effectively constructs a `merging tree' topology, where the fusion process propagates bottom-up and terminates once the budget constraint is satisfied. 
The pair formation logic for each layer adapts to the merging strategy: 

\begin{itemize}[leftmargin=1em]
    \item For Symmetric Merging, we target the weakest signals collectively. 
    The system sorts the chunks by relevance score, and partitions them into $\left\lfloor \frac{\left|\mathcal{C}^{(t)}\right|}{2}\right \rfloor$ disjoint pairs in a bottom-up manner:  
    the bottom two are paired, followed by the next two, \etc{}
    This strategy ensures maximum consolidation of weak signals, leaving the single highest-scoring chunk unmerged if the set size is odd.
    \item For Asymmetric Merging, we perform a greedy matching pass over the candidate pool:
    in each step, we pick the lowest-scoring available chunk, match it with its best semantic anchor with minimum NLL, and reserve this pair from the pool.
    This select-and-remove mechanism repeats until no valid source-anchor pairs remain, ensuring that each low-scoring chunk finds its most specific anchor without conflict.
\end{itemize}

\textbf{Analysis:}
Let $N$ denote the initial number of chunks, $M$ denote the final count, and $B$ denote the maximum concurrency limit (\ie{}, GPU batch size).
\begin{itemize}[leftmargin=1em]
    \item \textbf{Latency Efficiency:} 
    The sequential approach is strictly bound by $N-M$ serial operations. 
    In contrast, since the number of chunks approximately halves in each iteration, the hierarchical strategy exploits GPU parallelism to obtain only $\left\lceil \log\frac{N}{M}\right \rceil$ sequential steps when $N\leq 2B$.
    Even for $N>2B$ when the initial dense layers necessitate multiple batches, the total number of sequential inference rounds is bounded by
    \[
    \sum_{\frac{N}{2^k} > M} \max\left\{ \left\lceil \frac{N/2^k}{B} \right\rceil, 1\right\} \leq \left \lceil \frac{N}{B}+\log \frac{N}{M} \right \rceil, 
    \]
    which offers a significant acceleration compared to the linear $N-M$ cost of sequential merging once $B\geq 2$. 
    \item \textbf{Error Mitigation:}
    Sequential merging suffers from deep iterative dependency, where early-processed chunks may undergo up to $N-M$ rounds of regeneration, accumulating hallucinations and semantic drift.
    Hierarchical merging effectively flattens this dependency structure, capping the maximum generation depth at $\left\lceil \log \frac{N}{M} \right \rceil$.
    The reduction from linear to logarithmic significantly limits generation errors to preserve higher semantic fidelity.
\end{itemize}

%% file: content/4_experiments.tex
\section{Experiments}

\subsection{Experimental Setup}

\textbf{Datasets:}
We evaluate \ourname{} on five representative benchmarks.
Specifically, 2WikiMQA \cite{ho2020constructing}, HotpotQA \cite{yang2018hotpotqa}, and MuSiQue \cite{trivedi2022musique} are selected to evaluate the system's ability to retrieve and synthesize fragment evidence required for complex multi-hop reasoning.
TriviaQA \cite{joshi2017triviaqa} is used to challenge the ability to identify relevant signals in long-context scenarios. 
Finally, we include QASPER \cite{dasigi2021dataset} to assess the performance in finding answers within research papers.

\begin{table*}[!ht]
\centering
\caption{Experimental results of different methods across multiple datasets. The best and second-best results are highlighted in \textbf{bold} and \underline{underlined} respectively.}
\label{tab:rag-results}
\resizebox{\textwidth}{!}{%
\begin{tabular}{l|ccc|ccc|ccc|ccc|ccc}
\toprule
\multirow{2.5}{*}{Methods} & \multicolumn{3}{c|}{\makecell{2WikiMQA}} & \multicolumn{3}{c|}{HotpotQA} & \multicolumn{3}{c|}{MuSiQue} & \multicolumn{3}{c|}{TriviaQA} & \multicolumn{3}{c}{QASPER} \\
\cmidrule(lr){2-4} \cmidrule(lr){5-7} \cmidrule(lr){8-10} \cmidrule(lr){11-13} \cmidrule(lr){14-16} 
 & EM & F1 & Acc & EM & F1 & Acc & EM & F1 & Acc & EM & F1& Acc & EM & F1 & Acc \\
\midrule
BM25 & 27.5 & 35.1 & 34.5 & 35.5 & 46.7 & 42.5 & 9.5 & 16.5 & 15.0 & 23.5 & 44.9 & 31.0 & 9.5 & 23.0 & 10.5 \\
BGE-reranker & 32.5 & 42.3 & 43.5 & \underline{42.0} & \textbf{55.0} & 51.0 & 15.5 & 23.8 & 21.0 & 23.0 & \underline{45.6} & 31.0 & 10.0 & 31.0 & 13.0 \\
\midrule
RECOMP & 27.0 & 34.3 & 38.0 & 31.0 & 44.4 & 40.0 & 14.5 & 21.6 & 18.0 & 23.5 & 45.5 & 31.5 & 11.5 & 26.3 & 14.0 \\
\midrule
RAPTOR & 36.2 & 45.2 & 45.2 & 34.5 & 48.9 & 44.0 & 14.0 & 20.2 & 19.0 & \underline{24.2} & 45.4 & 31.8 & 11.5 & 30.9 & 13.5 \\
Tree-RAG & 22.5 & 31.8 & 32.5 & 27.5 & 36.2 & 30.8 & 5.0 & 14.9 & 15.0 & \textbf{25.0} & \textbf{48.1} & 31.0 & 7.5 & 17.5 & 8.1 \\
\midrule
\textbf{Ours-Sym} & \underline{41.5} & \underline{52.3} & \underline{55.5} & \textbf{43.5} & \textbf{55.0} & \underline{51.5} & \underline{23.0} & \underline{35.5} & \underline{30.0} & 24.0 & 44.9 & \textbf{32.5} & \textbf{15.0} & \textbf{38.4} & \underline{17.5} \\
\textbf{Ours-Asym} & \textbf{44.0} & \textbf{56.0} & \textbf{61.0} & 40.0 & \textbf{55.0} & \textbf{53.5} & \textbf{24.0} & \textbf{35.9} & \textbf{33.5} & 23.5 & 44.7 & \textbf{32.5} & \underline{12.5} & \underline{36.2} & \textbf{18.0} \\
\bottomrule
\end{tabular}
}
\end{table*}

\textbf{Baselines:}
To demonstrate the effectiveness of \ourname{}, we consider three major categories of RAG baselines. 
For selection-based solutions, we employ standard RAG with both sparse retriever BM25 \cite{robertson2009probabilistic} and dense retriever BGE-reranker-v2-M3 \cite{chen2024m3} to provide a comprehensive comparison. 
We also evaluate context compression techniques, specifically RECOMP \cite{xu2023recomp}, to examine performance in token-constrained scenarios. 
Finally, we compare our method with RAPTOR \cite{sarthi2024raptor} and Tree-RAG \cite{tao2025treerag}, which represent the knowledge aggregation paradigm by building query-agnostic hierarchical structures for comprehensive information synthesis.
Notably, we exclude iterative or adaptive frameworks such as Self-RAG \cite{asai2024self} and ChunkRAG \cite{singh2024chunkrag} from our direct baselines. 
These methods rely on multiple retrieval-generation cycles to enhance answer quality. 
In contrast, \ourname{} operates within a single-retrieval paradigm and is orthogonal to such approaches.
While iterative methods focus on the dynamic expansion of the search space, our work targets the optimization of context utilization --- specifically, how to structure and compress information in a given window without multi-step retrieval.

\textbf{Implementation:}
All experiments were conducted on a high-performance computing cluster equipped with 96 NVIDIA H20 Tensor Core GPUs, each featuring 141 GB of HBM3 memory. 
To ensure a fair comparison, we employ BGE-reranker-v2-M3 \cite{chen2024m3} as the default retriever for \ourname{}, reranking methods, and other embedding-based baselines; 
it is also used to compute the query-relevance scores $S(c_i; q)$ that guide Symmetric and Asymmetric Merging processes. 
We use Qwen3-30B-A3B-Instruct-2507 \cite{qwen3technicalreport} as the unified generator across all experiments to eliminate the influence of different language models.\footnote{
We also run experiments with different generation models. See Appendix \ref{sec:model} for more experimental results. 
} 
For top-$k$ selection solutions, we report results for $k=5$, and we constrain the maximum token length $L$ to 5 times the average chunk length for compression-based algorithms and \ourname{} . 
% Other methods without fixed $k$ constraints are evaluated using their default selection strategies. 

\textbf{Evaluation:}
Following standard practices in multi-hop reasoning tasks, we evaluate the performance of all methods using token-level Exact Match (EM), F1 score, and Accuracy metrics on different datasets. 
EM measures the strictest form of correctness, calculating the percentage of predictions that match the ground truth exactly. 
F1 score provides an assessment of partial correctness by computing the harmonic mean of token-level precision and recall between the generated response and the reference. 
Finally, Accuracy is used to evaluate the algorithm's success in capturing key information: generated answers with the ground truth appearing as a substring are considered as correct. 

% F1 assesses the token-level overlap, EM measures the percentage of predictions that match the ground truth exactly,  while Accuracy quantifies the ability of identifying the correct answer from long-context information. 

\subsection{Main Results}

The experimental results are summarized in Table \ref{tab:rag-results}. 
To provide a deeper understanding of \ourname{}'s performance, we examine the findings from two primary aspects:

\textbf{Comparison between \ourname{} and baselines:}
The results show that \ourname{} clearly outperforms existing baselines on all datasets. 
Specifically, on multi-hop datasets like 2WikiMQA, HotpotQA and MuSiQue, both variants of \ourname{} obtain the best performance among all methods. 
On the MuSiQue dataset, \ourname{-Asym} surpasses the strongest baseline BGE-reranker by a substantial margin of 12.1 points in F1 score; 
\ourname{} also outperforms the query-agnostic baseline RAPTOR by 15.7 points, demonstrating the superiority of our query-aware merging over static clustering.
On the document-intensive QASPER dataset, \ourname{} witnesses a 5.2--7.4 points improvement in F1 score compared to state-of-the-art methods. 
Although \ourname{} does not yield optimal F1 score on the TriviaQA dataset, it still achieves the best Accuracy and competitive EM and F1 performance.

\textbf{Comparison between \ourname{-Sym} and \ourname{-Asym}:}
We further compare the effectiveness of Symmetric versus Asymmetric Merging for \ourname{}. 
The results demonstrate that \ourname{-Asym} stands out particularly in multi-hop reasoning scenarios (e.g., 2WikiMQA, MuSiQue). 
Since \ourname{-Asym} aggressively eliminates semantic redundancy by removing repetitive information, it preserves valuable token space for distinct reasoning steps within the constrained context budget.
Nonetheless, for datasets requiring high recall from long-tail documents (e.g., QASPER, TriviaQA), \ourname{-Sym} stands out for its ability to capture tail information. 
Unlike standard RAG which directly discards such low-ranking chunks, \ourname{} merges them into new chunks to recover valid evidence that would otherwise be lost.

\begin{figure*}[!t]
	\centering
    \includegraphics[height=0.52cm]{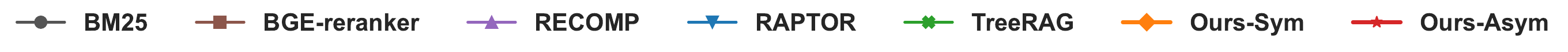}
    \begin{subfigure}[t]{0.2\textwidth}{
			\includegraphics[width=\textwidth]{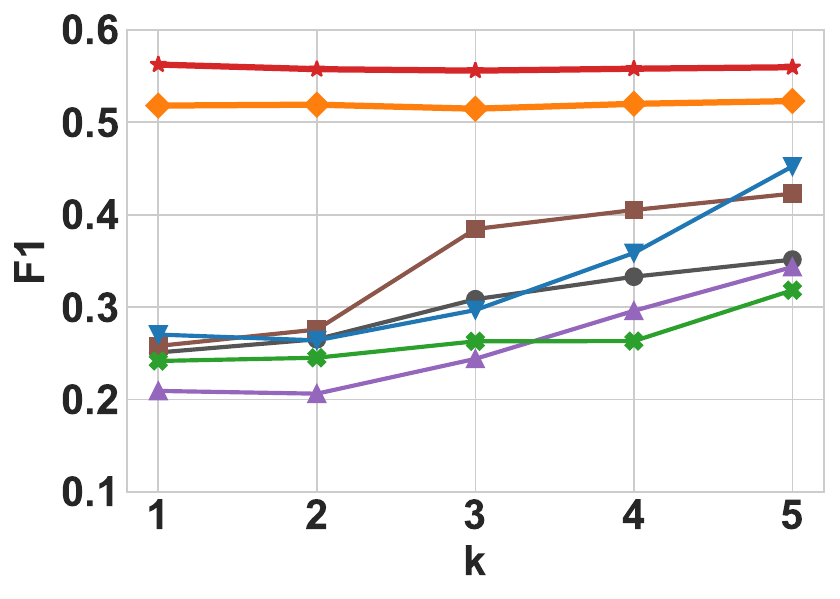}
            \caption{2WikiMQA}
		}
	\end{subfigure}%
	\begin{subfigure}[t]{0.2\textwidth}{
			\includegraphics[width=\textwidth]{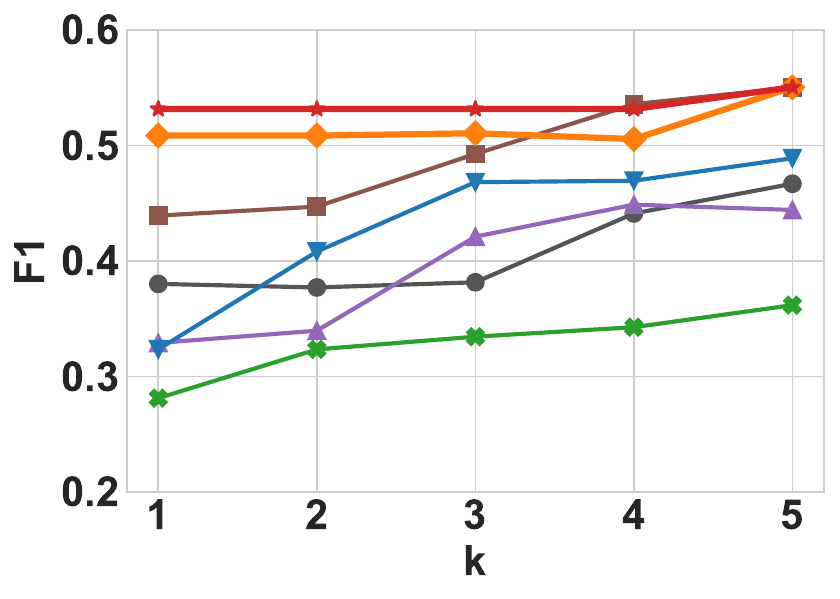}
            \caption{HotpotQA}
		}
	\end{subfigure}%
    \begin{subfigure}[t]{0.2\textwidth}{
			\includegraphics[width=\textwidth]{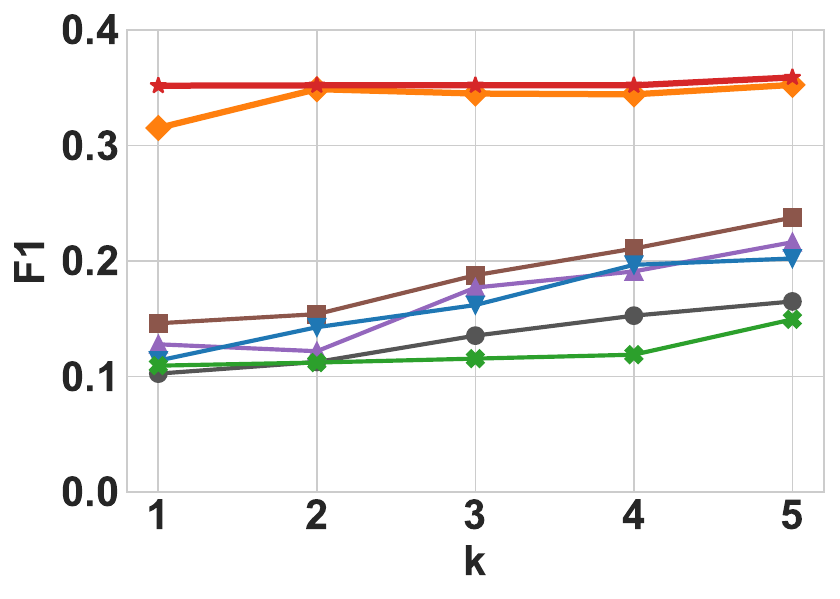}
            \caption{MuSiQue}
		}
	\end{subfigure}%
    \begin{subfigure}[t]{0.2\textwidth}{
			\includegraphics[width=\textwidth]{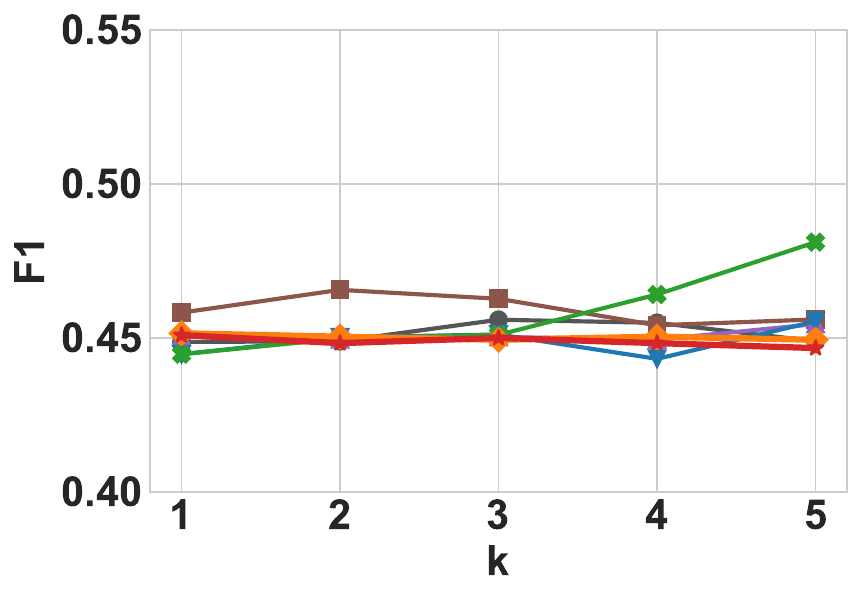}
            \caption{TriviaQA}
		}
	\end{subfigure}%
    \begin{subfigure}[t]{0.2\textwidth}{
			\includegraphics[width=\textwidth]{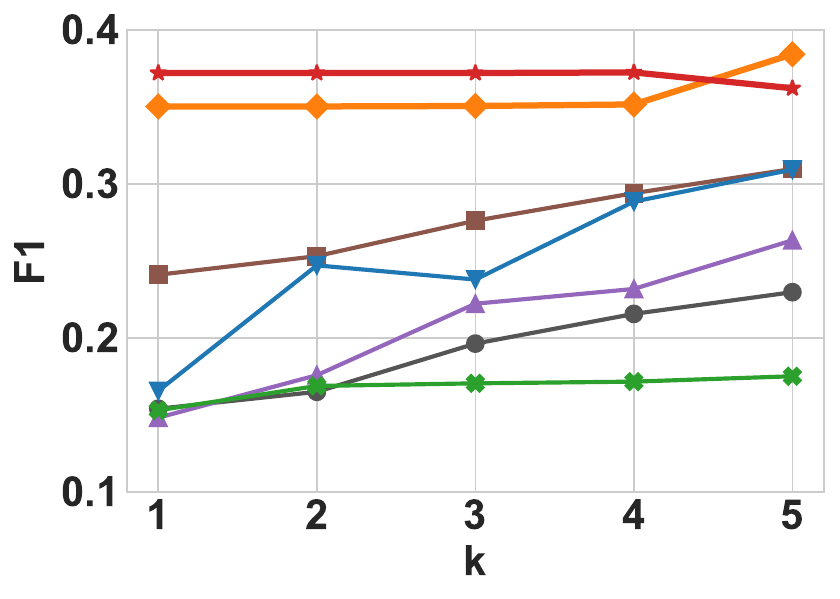}
            \caption{QASPER}
		}
	\end{subfigure}%
	\caption{Impact of different $k$ on F1.}
    \label{fig:f1-k}
\end{figure*}

\begin{figure*}[!t]
	\centering
    	\begin{subfigure}[t]{0.5\textwidth}{
			\includegraphics[width=\textwidth]{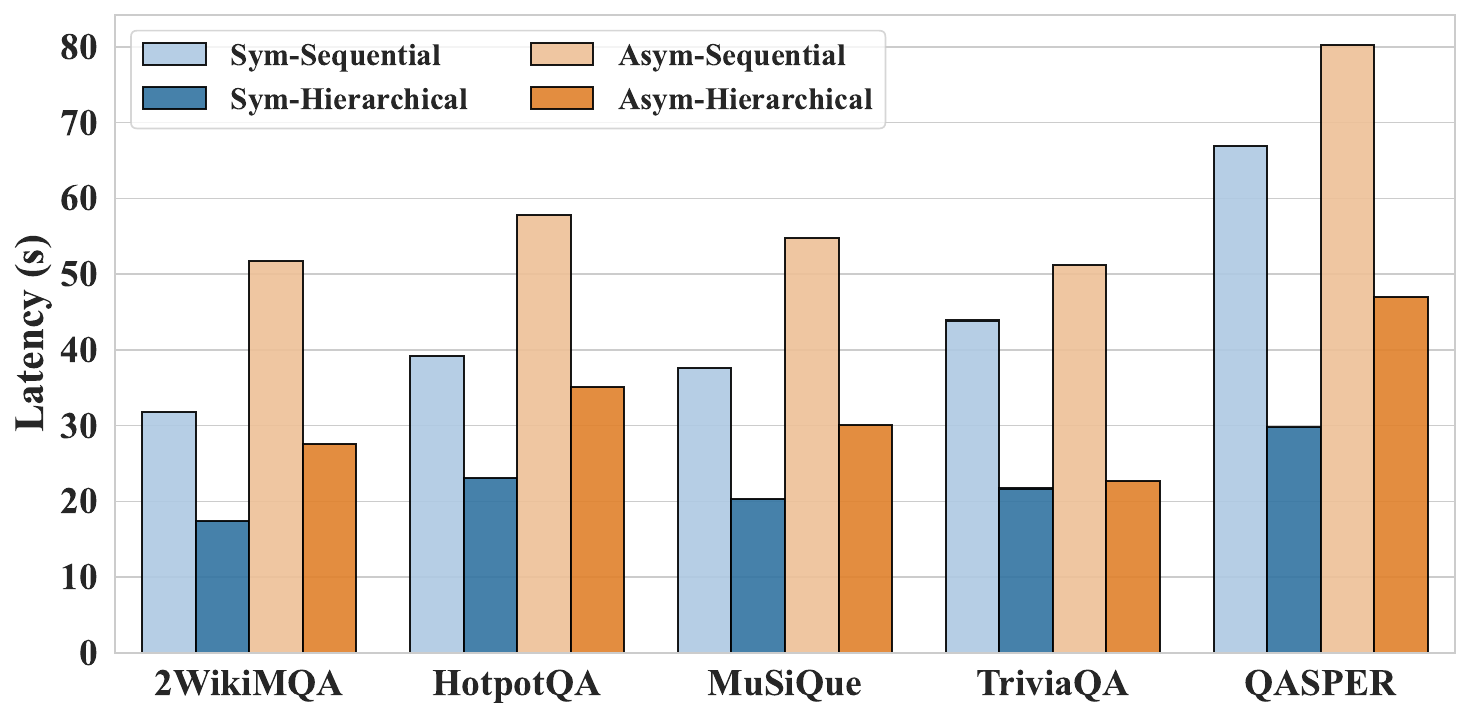}
			\label{fig:nll}
            \caption{Latency}
		}
	\end{subfigure}%
	\begin{subfigure}[t]{0.5\textwidth}{
			\includegraphics[width=\textwidth]{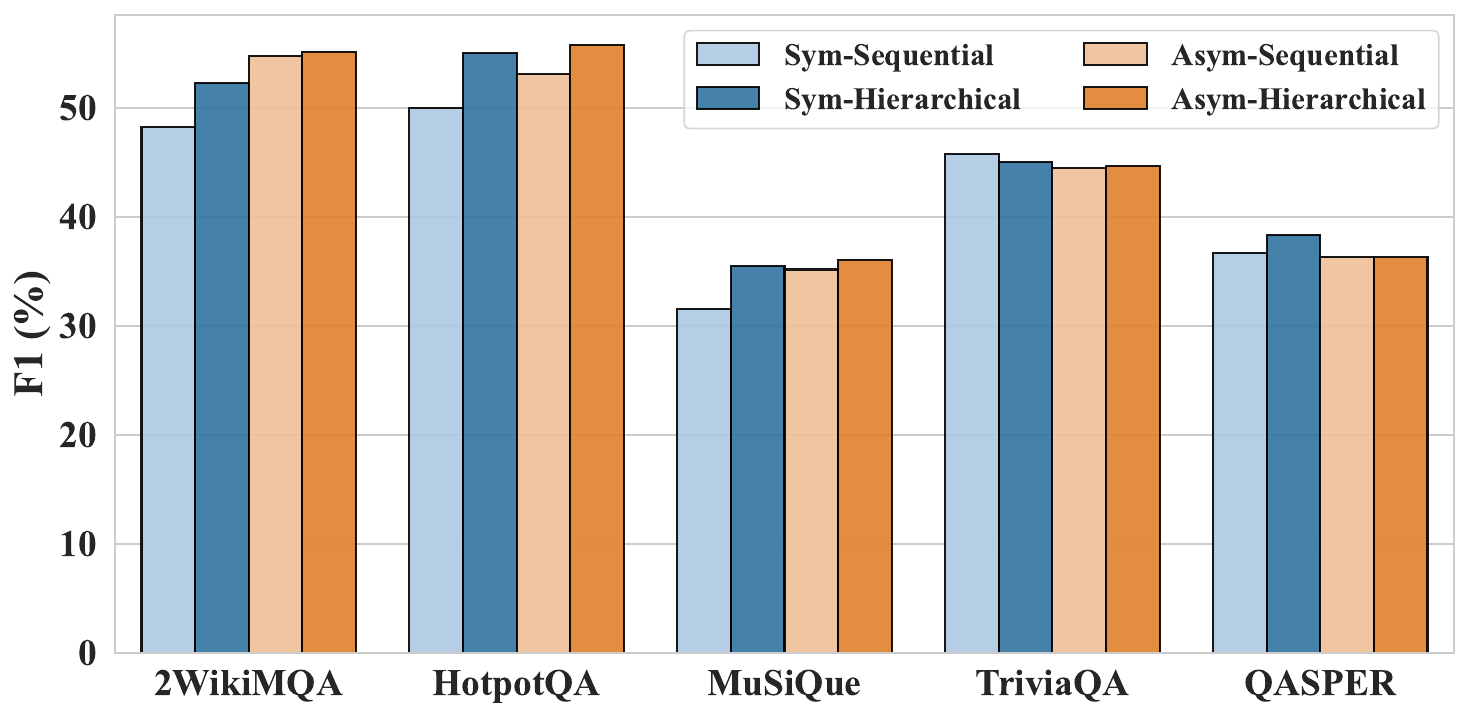}
			\label{fig:cos}
            \caption{F1}
		}
	\end{subfigure}
	\caption{Ablation on Hierarchical Parallel Merging }
    \label{fig:parallel}
\end{figure*}

\subsection{Impact of Context Length}

In this section, we investigate the impact of $k$ to demonstrate the advantage of \ourname{} within token-constrained scenarios.
Specifically, we vary $k$ from 1 to 5, and set the maximum context window budget $L$ to $k$ times the average chunk length. 
The F1 scores of \ourname{} and the baseline solutions are presented in Figure \ref{fig:f1-k}.\footnote{
We only report F1 score in the main text; the corresponding results for EM and Accuracy are provided in Appendix \ref{sec:top-appendix}. 
}

We observe that \ourname{} maintains exceptional robust performance even within limited context budget. 
At the extreme setting when $k=1$, \ie{}, the maximum context length is equal to merely a single chunk, \ourname{} successfully extracts and synthesizes useful information from multiple chunks to construct a high-quality context.
This result validates our core design philosophy: our method compresses redundant information to fit multi-source evidence into a minimal window, while selecting a single best chunk may be incomplete for RAG. 

In contrast, other baselines suffer from a sharp decline in F1 scores as $k$ decreases: 
Notably, when $k=1$, the F1 scores of baselines drop to approximately half of \ourname{}, highlighting their inability to effectively organize information across different chunks. 
When the answer requires reasoning across disjoint pieces of evidence, or when the top-1 chunk is noisy, their inability to merge information leads to retrieval failure.
Furthermore, although their performance gradually recovers as the constraint is relaxed, it remains consistently inferior to \ourname{}.

\subsection{Efficiency Analysis}

To empirically validate the scalability of our proposed Hierarchical Parallel Merging, we compare its performance against the sequential merging version of \ourname{}.
We use vLLM \cite{kwon2023efficient} as the inference backend, and set the parallel parameter $B=8$ for hierarchical merging. 
We measure the latency for each query to quantify the acceleration ratio, while simultaneously evaluating accuracy metrics (\ie{}, EM, F1, accuracy) to assess the balance between computational efficiency and synthesis quality in Figure \ref{fig:parallel}.\footnote{
We only report F1 score and average query latency in the main text; 
detailed results for EM and Accuracy are provided in Appendix \ref{sec:hierarchical-appendix}. 
}

\textbf{Effects on Latency:}
After implementing hierarchical strategy, the generation latency achieves up to 2.24$\times$/2.26$\times$ acceleration for Symmetric Merging and Asymmetric Merging respectively.
Although such speedup is sub-linear relative to the batch size ($B=8$), it is primarily attributed to two factors: 
\ding{172} the chunk selection process incurs additional latency, especially the computation of NLL which involves dense forward passes; 
\ding{173} vLLM itself already employs continuous batching and request scheduling optimization. 
Nevertheless, the $>2\times$ speedup still remains significant for real-time applications.

\textbf{Effects on F1 Score:}
The experimental results illustrate that the hierarchical strategy achieves comparable or even superior generation quality \wrt{} F1 score.
On the 2WikiMQA, HotpotQA and MuSiQue datasets, Hierarchical Batched Merging achieves at least a 4\% improvement for Symmetric Merging, and a 0.5\%--1.5\% gain for Asymmetric Merging. 
This strategy also brings marginal benefits to the QASPER dataset, with a 1.6\% increase for Symmetric Merging. 
Notably, while a slight dip in F1 score is observed on TriviaQA dataset compared to the sequential baseline, this degradation is negligible, which is less than 0.8\% for Symmetric Merging.

%% file: content/5_conclusion.tex
\section{Conclusion}

In this paper, we reconceptualize the retrieval-augmented generation problem as a dynamic context construction problem rather than a static chunk selection task. 
Addressing the inherent limitations of existing retrieve-then-select pipelines, we propose \ourname{}, a novel query-aware framework for RAG context synthesis. 
Specifically, \ourname{} applies \textbf{Symmetric Merging} to maintain information in the tail, and \textbf{Asymmetric Merging} to eliminate semantic redundancy. 
In addition, we introduce the \textbf{Hierarchical Parallel Merging} strategy to reduce information loss and accelerate execution speed during the merging operation. 
Extensive experiments across different benchmarks demonstrate the superiority of \ourname{} over state-of-the-art algorithms, validating the efficacy of retrieval-as-synthesis.

%% file: content/6_appendix.tex
\section{Pseudo-code of \ourname{}}

\subsection{Symmetric and Asymmetric Merging}

The pseudo-code of the merging operation is shown in Algorithm \ref{alg:sym} (symmetric) and Algorithm \ref{alg:asym} (asymmetric).

\begin{algorithm}[H]
    \caption{Symmetric Merging Operation}
    \label{alg:sym}
    \begin{algorithmic}
        \STATE {\textbf{Input:}} Set of chunks $\mathcal{C} = \{c_1, c_2, \cdots, c_N\}$, user query $q$, maximum context length $L$
        \STATE $t\leftarrow 0$, $\mathcal{C}^{(0)}\leftarrow \mathcal{C}$
        \WHILE{$\underset{c\in \mathcal{C}^{(t)}}{\sum} l(c) > L$ \AND $\left|\mathcal{C}^{(t)}\right| > 1$}
            \STATE $(c_i^*, c_j^*) \leftarrow \underset{c_i, c_j \in \mathcal{C}^{(t)}, i\neq j}{\arg \min} \left\{S(c_i;q)+ S(c_j;q)\right\}$
            \STATE $c^* \leftarrow \mathcal{M}(c_i^*, c_j^*; q)$
            \STATE $\mathcal{C}^{(t+1)} \leftarrow \left(\mathcal{C}^{(t)} \setminus \{c_i^*, c_j^*\}\right) \cup \{c^*\}$
            \STATE $t\leftarrow t+1$
        \ENDWHILE
        \STATE $\mathcal{S}\leftarrow \mathcal{C}^{(t)}$
        \RETURN $\mathcal{S}$
    \end{algorithmic}
\end{algorithm}

\begin{algorithm}[H]
    \caption{Asymmetric Merging Operation}
    \label{alg:asym}
    \begin{algorithmic}
        \STATE {\textbf{Input:}} Set of chunks $\mathcal{C} = \{c_1, c_2, \cdots, c_N\}$, user query $q$, maximum context length $L$
        \STATE $t\leftarrow 0$, $\mathcal{C}^{(0)}\leftarrow \mathcal{C}$
        \WHILE{$\underset{c\in \mathcal{C}^{(t)}}{\sum} l(c) > L$ \AND $\left|\mathcal{C}^{(t)}\right| > 1$}
            \STATE $c_i^* \leftarrow \underset{c_i \in \mathcal{C}^{(t)}}{\arg\min} S(c_i; q)$
            \STATE $c_j^* \leftarrow \underset{c_j \in \mathcal{C}^{(t)}, j\neq i}{\arg\min}\mathcal{L}_{\text{NLL}}(c_i^*|c_j)$
            \STATE $c^* \leftarrow \mathcal{M}(c_i^*, c_j^*; q)$
            \STATE $\mathcal{C}^{(t+1)} \leftarrow \left(\mathcal{C}^{(t)} \setminus \{c_i^*, c_j^*\}\right) \cup \{c^*\}$
            \STATE $t\leftarrow t+1$
        \ENDWHILE
        \STATE $\mathcal{S}\leftarrow \mathcal{C}^{(t)}$
        \RETURN $\mathcal{S}$
    \end{algorithmic}
\end{algorithm}

\subsection{Hierarchical Parallel Merging}

The pseudo-code for Hierarchical Parallel Merging is shown in Algorithm \ref{alg:sym:par} (symmetric) and Algorithm \ref{alg:asym:par} (asymmetric).

\begin{algorithm}[H]
    \caption{Batched Symmetric Merging Operation}
    \label{alg:sym:par}
    \begin{algorithmic}
        \STATE {\textbf{Input:}} Set of chunks $\mathcal{C} = \{c_1, c_2, \cdots, c_N\}$, user query $q$, maximum context length $L$
        \STATE $t\leftarrow 0$, $\mathcal{C}^{(0)}\leftarrow \mathcal{C}$
        \WHILE{$\underset{c\in \mathcal{C}^{(t)}}{\sum} l(c) > L$ \AND $\left|\mathcal{C}^{(t)}\right| > 1$}
            \STATE $\mathcal{R} \leftarrow \mathcal{C}^{(t)}$ \quad \texttt{// Temporary remaining set}
            \STATE $\mathcal{P} \leftarrow \varnothing$ \quad \texttt{// Set of pairs to merge}
            \STATE $K \leftarrow \left\lfloor \frac{\left|\mathcal{C}^{(t)}\right|}{2}\right\rfloor$ \quad \texttt{// Adjust batch size}
            \FOR{$1\leq k \leq K$}
                \STATE $(c_{i}^*, c_{j}^*) \leftarrow \underset{c_i, c_j \in \mathcal{R}, i\neq j}{\arg\min} \left\{S(c_i; q)+S(c_j; q)\right\}$
                \STATE add $(c_{i}^*, c_{j}^*)$ to $\mathcal{P}$
                \STATE $\mathcal{R} \leftarrow \mathcal{R} \setminus \{c_{i}^*, c_{j}^*\}$
            \ENDFOR
            \STATE $\mathcal{C}_{new} \leftarrow \varnothing$
            \FOR{\textbf{each} $(c_i^*, c_j^*) \in \mathcal{P}$ \textbf{in parallel}}
                \STATE $c^* \leftarrow \mathcal{M}(c_i^*, c_j^*; q)$
                \STATE $\mathcal{C}_{new} \leftarrow \mathcal{C}_{new} \cup \{c^*\}$
            \ENDFOR
            \STATE $\mathcal{C}^{(t+1)} \leftarrow \mathcal{R} \cup \mathcal{C}_{new}$
            \STATE $t\leftarrow t+1$
        \ENDWHILE
        \STATE $\mathcal{S}\leftarrow \mathcal{C}^{(t)}$
        \RETURN $\mathcal{S}$
    \end{algorithmic}
\end{algorithm}

\begin{algorithm}[H]
    \caption{Batched Asymmetric Merging Operation}
    \label{alg:asym:par}
    \begin{algorithmic}
        \STATE {\textbf{Input:}} Set of chunks $\mathcal{C} = \{c_1, c_2, \cdots, c_N\}$, user query $q$, maximum context length $L$
        \STATE $t\leftarrow 0$
        \STATE $\mathcal{C}^{(0)}\leftarrow \mathcal{C}$
        \WHILE{$\underset{c\in \mathcal{C}^{(t)}}{\sum} l(c) > L$ \AND $\left|\mathcal{C}^{(t)}\right| > 1$}
            \STATE $\mathcal{R} \leftarrow \mathcal{C}^{(t)}$ \quad \texttt{// Temporary remaining set}
            \STATE $\mathcal{P} \leftarrow \varnothing$ \quad \texttt{// Set of pairs to merge}
            \STATE $K \leftarrow \left\lfloor \frac{\left|\mathcal{C}^{(t)}\right|}{2}\right\rfloor $ \quad \texttt{// Adjust batch size}
            \FOR{$1\leq k \leq K$}
                \STATE $c_{i}^* \leftarrow \underset{c_i \in \mathcal{R}}{\arg\min} S(c_i; q)$
                \STATE $c_{j}^* \leftarrow \underset{c_j \in \mathcal{R}, j\neq i}{\arg\min}\mathcal{L}_{\text{NLL}}(c_{i}^*|c_j)$
                \STATE add $(c_{i}^*, c_{j}^*)$ to $\mathcal{P}$
                \STATE $\mathcal{R} \leftarrow \mathcal{R} \setminus \{c_{i}^*, c_{j}^*\}$
            \ENDFOR
            \STATE $\mathcal{C}_{new} \leftarrow \varnothing$
            \FOR{\textbf{each} $(c_i^*, c_j^*) \in \mathcal{P}$ \textbf{in parallel}}
                \STATE $c^* \leftarrow \mathcal{M}(c_i^*, c_j^*; q)$
                \STATE $\mathcal{C}_{new} \leftarrow \mathcal{C}_{new} \cup \{c^*\}$
            \ENDFOR
            \STATE $\mathcal{C}^{(t+1)} \leftarrow \mathcal{R} \cup \mathcal{C}_{new}$
            \STATE $t\leftarrow t+1$
        \ENDWHILE
        \STATE $\mathcal{S}\leftarrow \mathcal{C}^{(t)}$
        \RETURN $\mathcal{S}$
    \end{algorithmic}
\end{algorithm}

\section{Prompts for Merging and Generation}

We design prompts that translate the theoretical objectives of \ourname{} into executable LLM instructions. 
Specifically, the prompt for Symmetric Merging is shown as follows:

\begin{promptbox}{Prompt for Symmetric Merging}

\textbf{Task:} 

Merge Chunk A and Chunk B into a single context based on the Query. 

\textbf{Instructions:}

\begin{itemize}
    \item \textbf{Selective Retention:} 
    Extract and keep all relevant sentences or phrases from Chunk A and Chunk B that address the Query verbatim. 
    Do not paraphrase or modify the original text. 
    \item \textbf{Pruning:} 
    Remove any content that is irrelevant to the Query. 
    \item \textbf{No Generation:}
    Do not answer the Query. 
    Do not generate any new information or external knowledge. 
    \item \textbf{Output:} 
    Provide only the merged verbatim text. 
\end{itemize}

\textbf{Input Data:} 

Query: \{\textit{query}\} 

Chunk A : \{\textit{chunk\_a}\} 

Chunk B: \{\textit{chunk\_b}\} 

\textbf{Output:}

\end{promptbox}

The prompt for Asymmetric Merging is shown as follows:

\begin{promptbox}{Prompt for Asymmetric Merging}
\textbf{Task:}

Construct a merged context by using Chunk A as the primary anchor and filling in missing details from Chunk B. 

\textbf{Instructions:}

\begin{itemize}
    \item \textbf{Role Definition:} Chunk A is the Anchor (Primary Context). Chunk B is Supplementary Material.
    \item \textbf{Strict Verbatim Policy:} You must extract exact sentences or phrases from the original text. Do not rewrite, paraphrase, or modify any words from either chunk.
    \item \textbf{Processing Chunk A (Anchor):}
    Retain all sentences in Chunk A that are relevant to the Query.
    Remove only the parts of Chunk A that are completely irrelevant.
    \item \textbf{Processing Chunk B (Supplement):}
    Extract sentences from Chunk B only if they meet two conditions:
    \begin{itemize}
        \item Strictly relevant to the Query.
        \item Contains information NOT present in Chunk A (Deduplication).
    \end{itemize}
    \item \textbf{Assembly:} 
    Combine the retained parts of Chunk A and the selected unique parts of Chunk B into a single context. 
    Maintain the logical flow of Chunk A as the baseline.
    \item \textbf{No Generation:} Do not answer the Query. Do not add connecting words. Do not generate external knowledge.
    \item \textbf{Output:} Provide only the merged verbatim text.
\end{itemize}
\textbf{Input Data:} 

Query: \{\textit{query}\} 

Chunk A (Anchor): \{\textit{chunk\_a}\} 

Chunk B (Source): \{\textit{chunk\_b}\} 

\textbf{Output:}

\end{promptbox}

The prompt for generating the final answer is shown as follows:

\begin{promptbox}{Prompt for Generation}
Answer the question based ONLY on the provided context.

\textbf{Instructions:}

\begin{enumerate}
    \item \textbf{Directness}: Provide the answer immediately without preamble.
    \item \textbf{Prioritization}: If the context mentions a primary method/entity along with minor exceptions or auxiliary steps, provide ONLY the primary one.
    \item \textbf{Ambiguity}: If multiple candidates exist (e.g., multiple baselines), use the specific details in the question to select the correct one.
    \item \textbf{Format}: If the answer is a list, provide only the key items. If it's a Yes/No question, output ONLY `Yes' or `No'.
\end{enumerate}

Context:
\{\textit{context}\}

Question: \{\textit{question}\}

Answer:

\end{promptbox}

\section{Detailed Metrics for Different $k$}
\label{sec:top-appendix}

\begin{figure*}[!ht]
	\centering
    \includegraphics[height=0.52cm]{figure/shared_legend_1row.pdf}
    \begin{subfigure}[t]{0.2\textwidth}{
			\includegraphics[width=\textwidth]{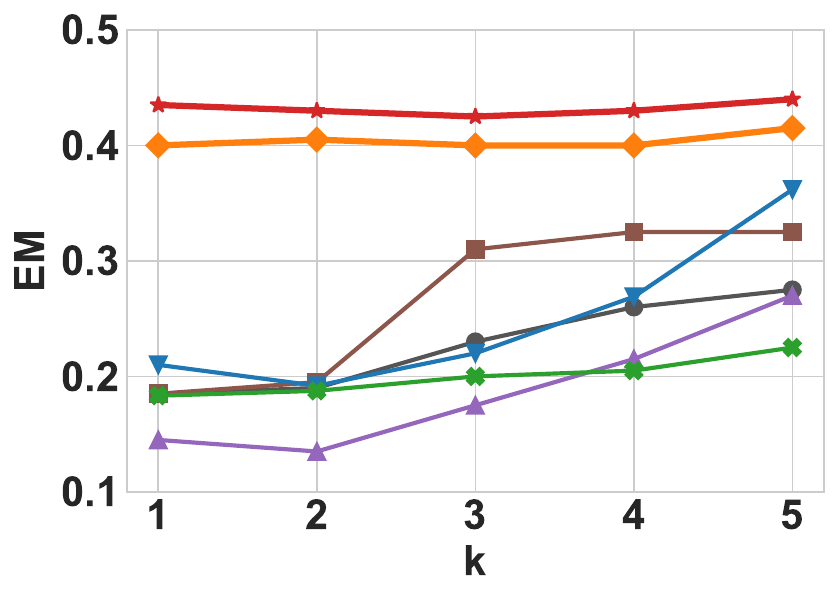}
            \caption{2WikiMQA}
		}
	\end{subfigure}%
	\begin{subfigure}[t]{0.2\textwidth}{
			\includegraphics[width=\textwidth]{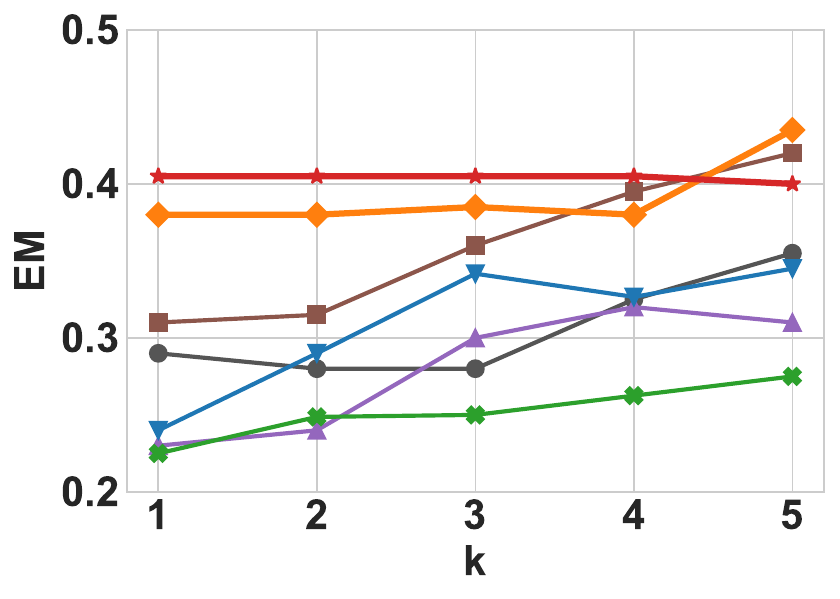}
            \caption{HotpotQA}
		}
	\end{subfigure}%
    \begin{subfigure}[t]{0.2\textwidth}{
			\includegraphics[width=\textwidth]{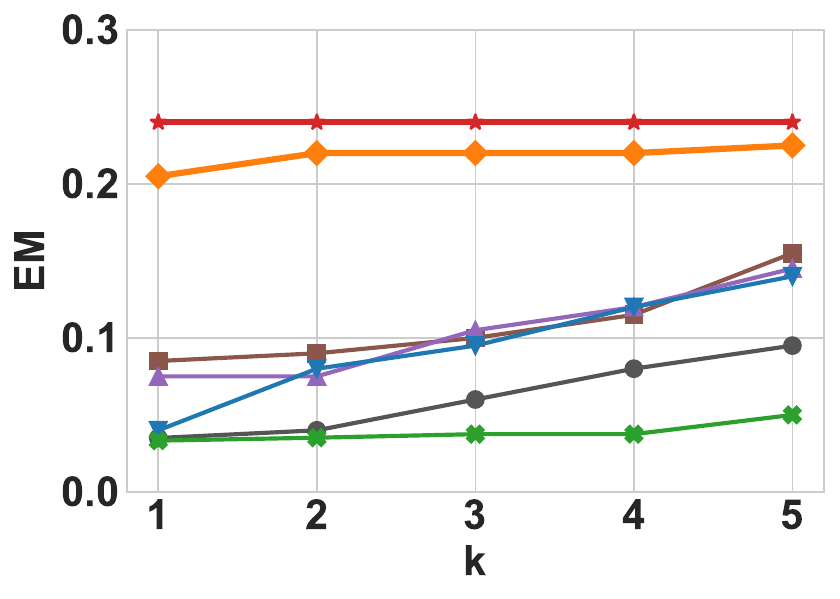}
            \caption{MuSiQue}
		}
	\end{subfigure}%
    \begin{subfigure}[t]{0.2\textwidth}{
			\includegraphics[width=\textwidth]{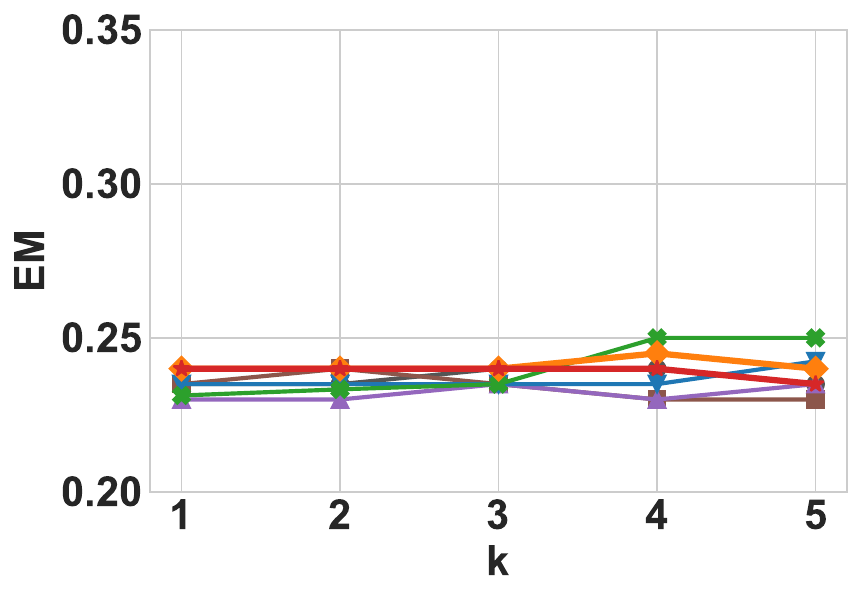}
            \caption{TriviaQA}
		}
	\end{subfigure}%
    \begin{subfigure}[t]{0.2\textwidth}{
			\includegraphics[width=\textwidth]{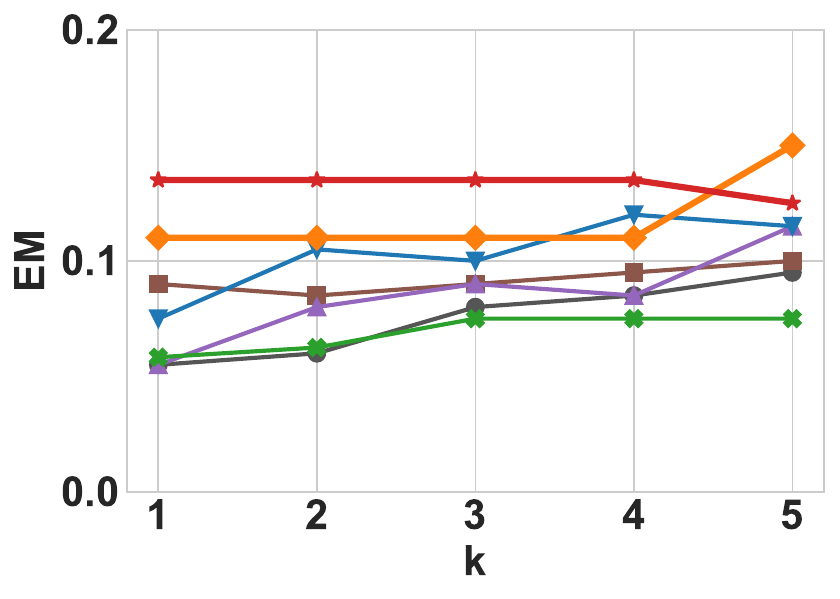}
            \caption{QASPER}
		}
	\end{subfigure}%
	\caption{Impact of different $k$ on EM.}
    \label{fig:em-k}
\end{figure*}

\begin{figure*}[!ht]
	\centering
    \includegraphics[height=0.52cm]{figure/shared_legend_1row.pdf}
    \begin{subfigure}[t]{0.2\textwidth}{
			\includegraphics[width=\textwidth]{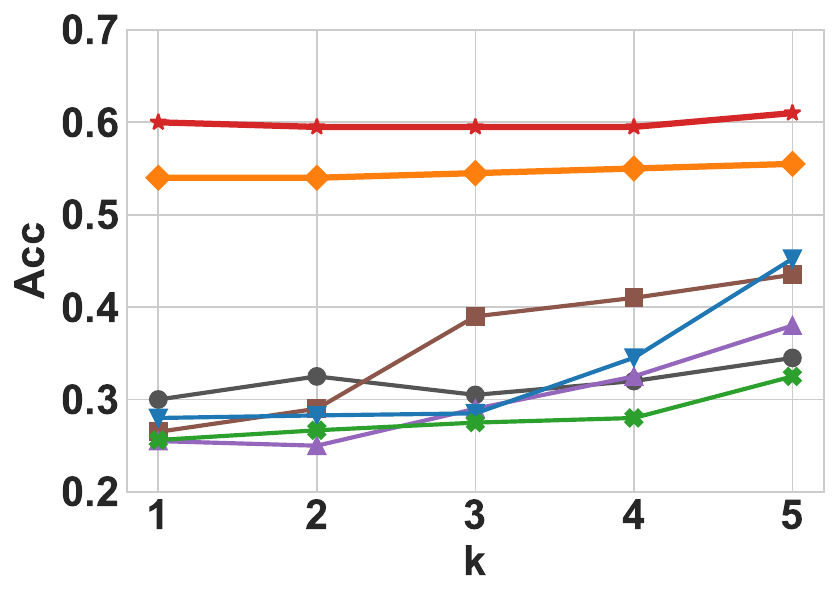}
            \caption{2WikiMQA}
		}
	\end{subfigure}%
	\begin{subfigure}[t]{0.2\textwidth}{
			\includegraphics[width=\textwidth]{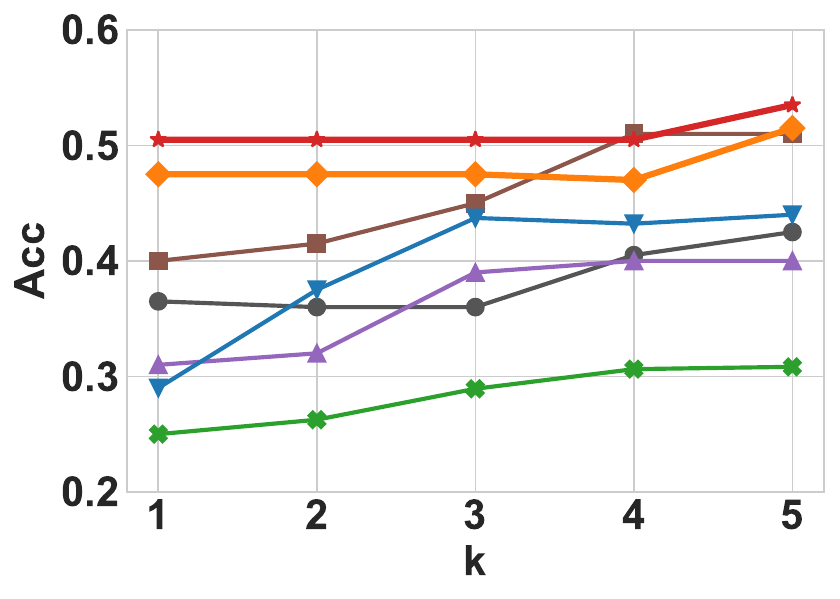}
            \caption{HotpotQA}
		}
	\end{subfigure}%
    \begin{subfigure}[t]{0.2\textwidth}{
			\includegraphics[width=\textwidth]{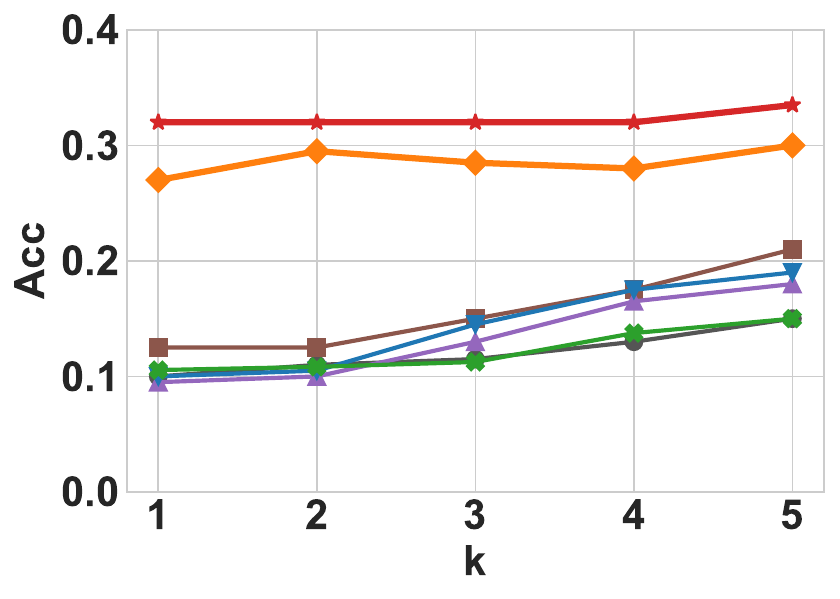}
            \caption{MuSiQue}
		}
	\end{subfigure}%
    \begin{subfigure}[t]{0.2\textwidth}{
			\includegraphics[width=\textwidth]{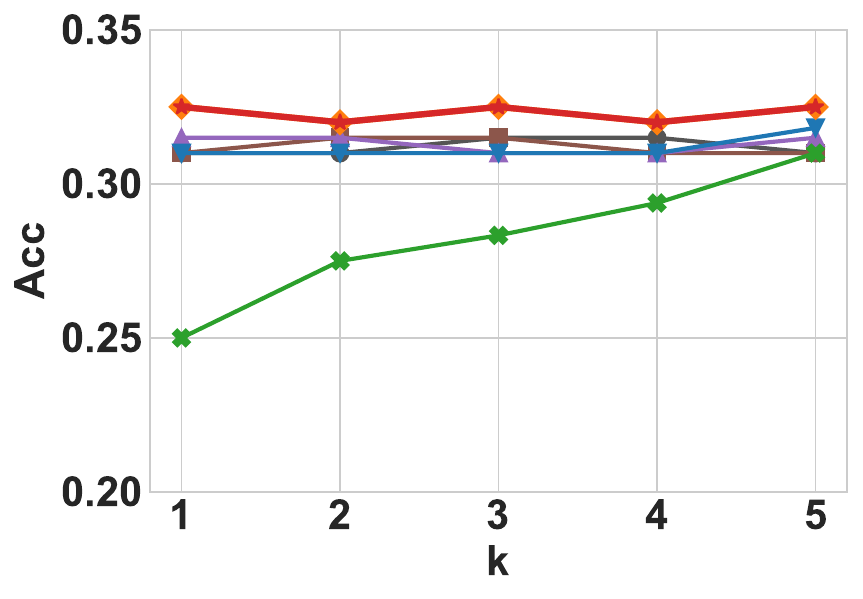}
            \caption{TriviaQA}
		}
	\end{subfigure}%
    \begin{subfigure}[t]{0.2\textwidth}{
			\includegraphics[width=\textwidth]{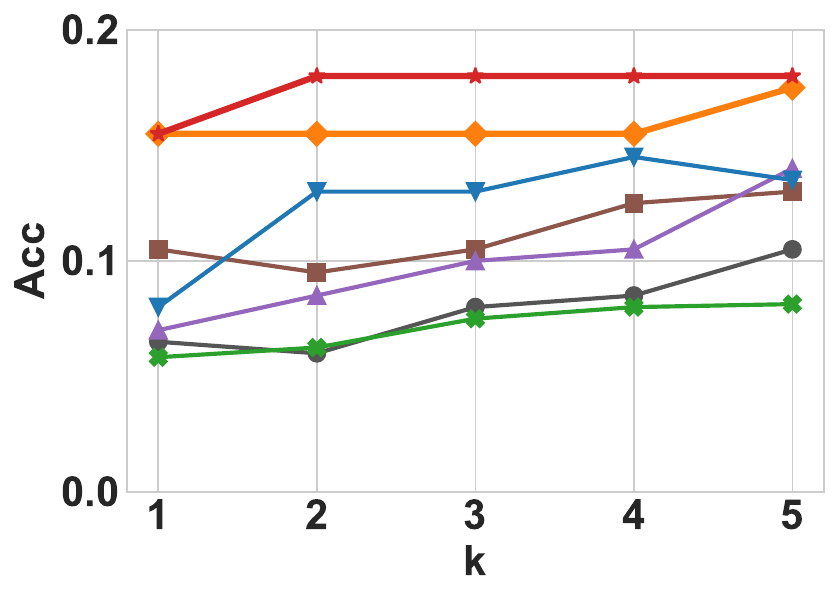}
            \caption{QASPER}
		}
	\end{subfigure}%
	\caption{Impact of different $k$ on Accuracy.}
    \label{fig:acc-k}
\end{figure*}

Here, we provide EM and Accuracy metrics for \ourname{} and various baselines. 
As illustrated in Figure \ref{fig:em-k} and \ref{fig:acc-k}, both metrics exhibit similar trends as $k$ decreases. 
While certain baselines achieve performance competitive with \ourname{} for larger $k$, their generation quality deteriorates sharply once the context length is reduced. 
On the 2WikiMQA dataset, the EM scores drop from $\sim$30\% ($k=5$) to $\sim$20\% ($k=1$), and Accuracy drops from up to 45\% ($k=5$) to 30\% ($k=1$) for state-of-the-art baselines. 
However, \ourname{} still achieves comparable generation quality using only a small fraction of the tokens. 
The EM score and Accuracy only show a marginal decrease of $\sim$5\% for \ourname{}, validating that it optimizes information density rather than relying on information redundancy.

\section{Detailed Metrics for Hierarchical Merging}
\label{sec:hierarchical-appendix}

\begin{figure*}[!ht]
	\centering
	\begin{subfigure}[t]{0.5\textwidth}{
			\includegraphics[width=\textwidth]{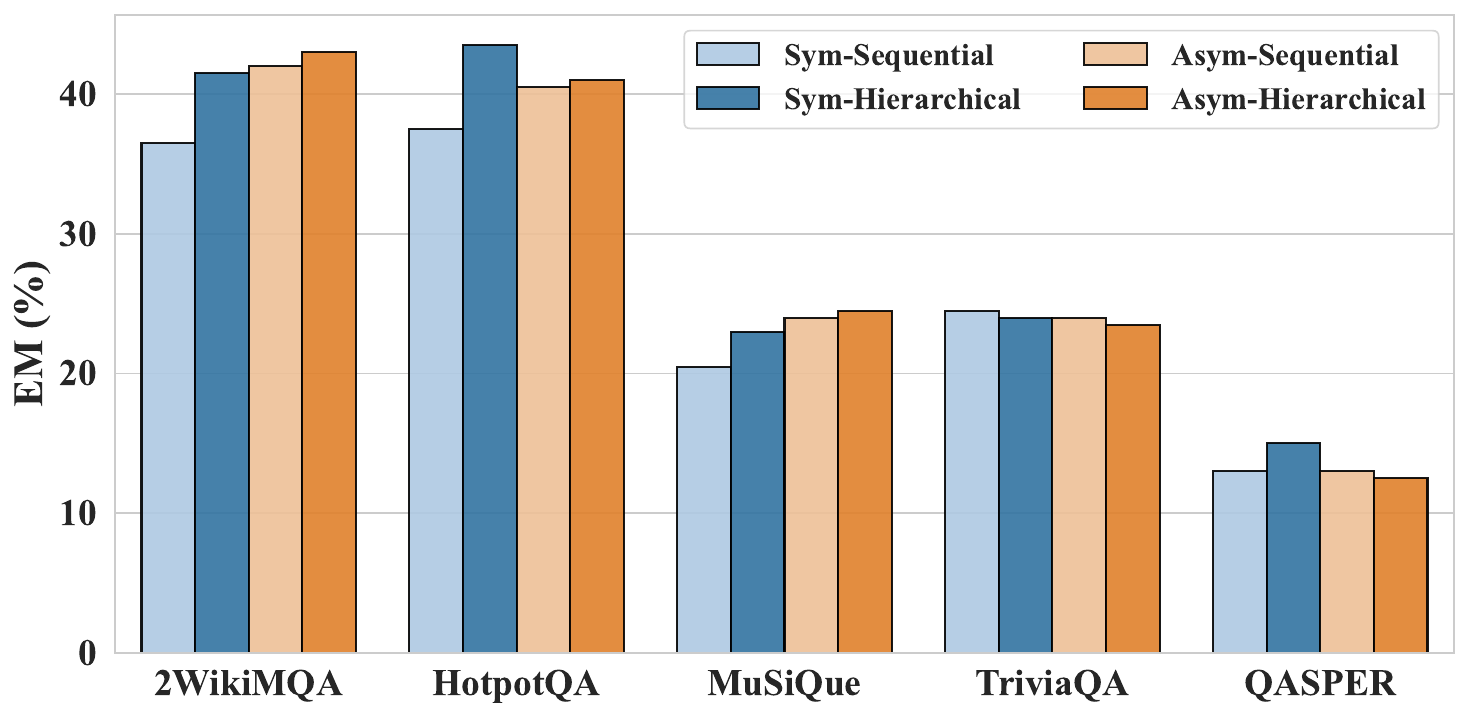}
			\label{fig:cos}
            \caption{EM}
		}
	\end{subfigure}%
	\begin{subfigure}[t]{0.5\textwidth}{
			\includegraphics[width=\textwidth]{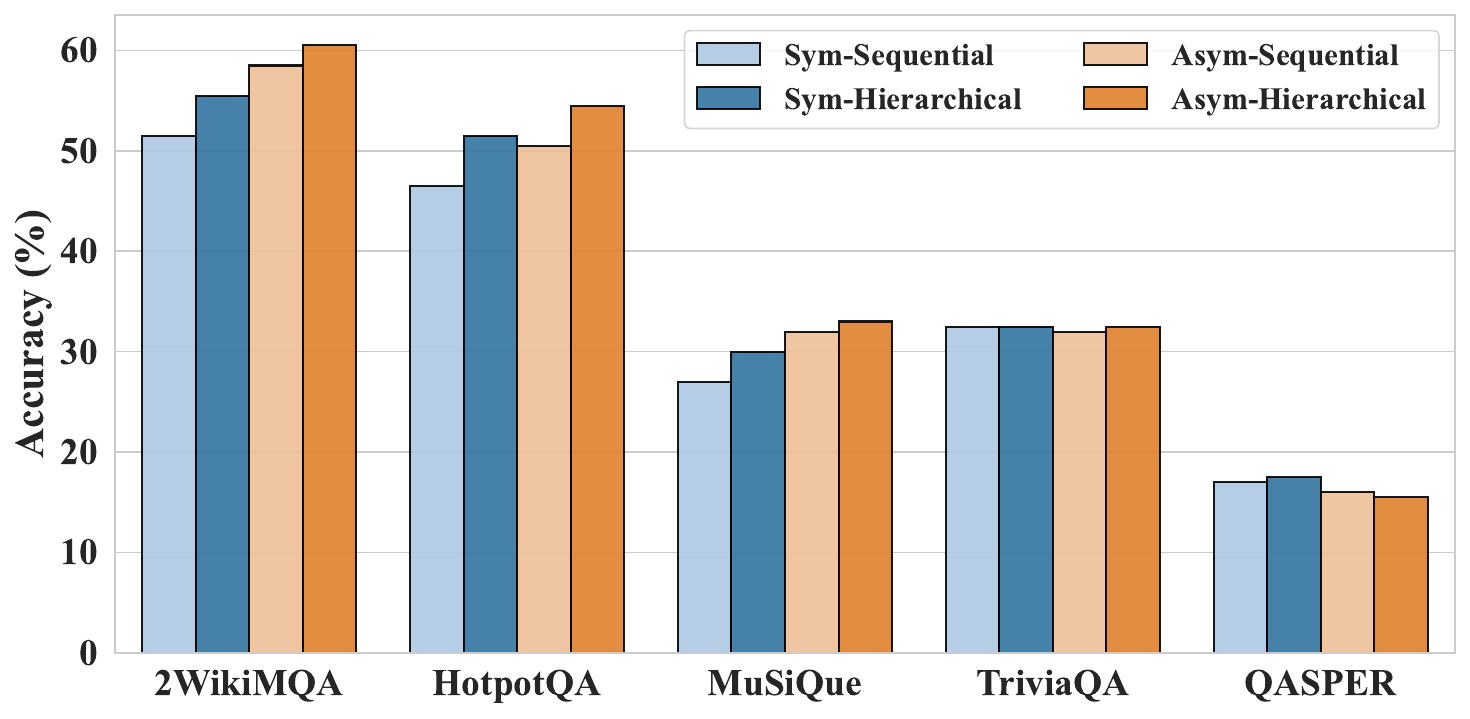}
			\label{fig:nll}
            \caption{Accuracy}
		}
	\end{subfigure}
	\caption{Ablation on Hierarchical Parallel Merging }
    \label{fig:parallel-more}
\end{figure*}

Here, we provide EM and Accuracy metrics for sequential and hierarchical pipelines of \ourname{}. 
As shown in Figure \ref{fig:parallel-more}, Hierarchical Parallel Merging generally achieves higher EM and Accuracy on all datasets. 
On the 2WikiMQA dataset, Hierarchical Parallel Merging achieves a 5\% increase in EM for Symmetric Merging, and a 1\% increase in EM for Asymmetric Merging. 
The Accuracy is also improved by 4\%/2\% for Symmetric/Asymmetric Merging on the 2WikiMQA dataset.
This positive trend extends to other multi-hop datasets such as HotpotQA and MuSiQue, where the hierarchical structure consistently yields higher EM and Accuracy scores compared to the sequential approach.
On the remaining datasets (TriviaQA and QASPER), the hierarchical approach maintains performance parity with negligible fluctuations, demonstrating that \ourname{} can achieve RAG acceleration without sacrificing generation quality.

\section{Model Generalizability Analysis}

\label{sec:model}

To demonstrate the generalizability of the \ourname{} framework, we run experiments with different backend models. 
Specifically, we also use Llama-3.1-8B-Instruct \cite{grattafiori2024llama} as the unified generator. 
Table \ref{tab:llama} summarizes the comparative results with the context budget fixed at $k=5$. 
The experimental results show that \ourname{} still maintains superior performance independently of the generation model: 
\ourname{} obtains optimal F1 score on the 2WikiMQA, MusiQue and QASPER datasets. 
Even on the HotpotQA and TriviaQA datasets where \ourname{} does not rank first, it achieves runner-up performance with margins negligible to the top baseline, validating its consistent effectiveness across different reasoning tasks.

\begin{table*}[!ht]
\centering
\caption{Experimental results on Llama-3.1-8B-Instruct. The best and second-best results are highlighted in \textbf{bold} and \underline{underlined} respectively.}
\label{tab:llama}
\resizebox{\textwidth}{!}{%
\begin{tabular}{l|ccc|ccc|ccc|ccc|ccc}
\toprule
\multirow{2.5}{*}{Methods} & \multicolumn{3}{c|}{\makecell{2WikiMQA}} & \multicolumn{3}{c|}{HotpotQA} & \multicolumn{3}{c|}{MuSiQue} & \multicolumn{3}{c|}{TriviaQA} & \multicolumn{3}{c}{QASPER} \\
\cmidrule(lr){2-4} \cmidrule(lr){5-7} \cmidrule(lr){8-10} \cmidrule(lr){11-13} \cmidrule(lr){14-16} 
 & EM & F1 & Acc & EM & F1 & Acc & EM & F1 & Acc & EM & F1& Acc & EM & F1 & Acc \\
\midrule
BM25 & 15.5 & 23.4 & 24.0 & 28.5 & 38.7 & 34.5 & 10.0 & 17.2 & 10.5 & 20.5 & 44.0 & 29.0 & 9.0 & 23.1 & 9.5 \\
BGE-reranker & 23.5 & \underline{33.5} & \textbf{35.5} & 27.5 & 36.2 & 30.8 & 14.5 & 22.9 & 16.5 & \textbf{22.5} & \textbf{46.0} & 30.0 & 10.5 & \underline{28.0} & 13.0 \\
\midrule
RECOMP & 18.0 & 24.2 & 24.5 & 27.5 & 40.0 & 32.0 & 10.5 & 15.3 & 11.0 & 21.0 & 44.6 & 29.7 & \textbf{12.6} & 27.9 & \underline{14.6} \\
\midrule
RAPTOR & 21.0 & 28.6 & 27.0 & 32.0 & \textbf{45.2} & \textbf{41.0} & 11.5 & 19.4 & 13.0 & 22.1 & 44.7 & 29.7 & \textbf{12.6} & 27.9 & \underline{14.6} \\
Tree-RAG & 17.0 & 21.8 & 24.0 & 18.0 & 24.9 & 24.0 & 2.0 & 5.1 & 2.5 & 20.0 & 43.5 & 29.5 & 3.0 & 5.0 & 3.0 \\
\midrule
\textbf{Ours-Sym} & \underline{25.0} & 33.4 & 34.5 & \textbf{33.5} & 42.4 & 39.0 & \underline{16.5} & \textbf{26.0} & \underline{20.0} & \textbf{22.5} & 45.2 & \underline{30.5} & 12.0 & \textbf{31.9} & \textbf{16.0} \\
\textbf{Ours-Asym} & \textbf{26.0} & \textbf{34.4} & \textbf{35.5} & \underline{32.5} & \underline{44.1} & \underline{40.5} & \textbf{19.0} & \underline{25.4} & \textbf{21.0} & \textbf{22.5} & \underline{45.7} & \textbf{31.0} & 11.5 & 25.7 & 13.5 \\
\bottomrule
\end{tabular}
}
\end{table*}

\section{A Qualitative Analysis Example}

Here, we present a representative example from the MuSiQue dataset to demonstrate the effectiveness of \ourname{}.
The query, \underline{When did military instruction start at the place where Larry Alcala was educated?}, necessitates a multi-hop reasoning chain: the system must first identify the institution where Alcala studied (Entity \ding{182}), and subsequently determine when military instruction began at that location (Target \ding{183}).

Initially, the context consists of Chunk A (85 tokens) with biographical details about Alcala, and Chunk B (678 tokens) containing historical records that might be relevant to the institution. 
When processed by \ourname{}, these fragments are fused into a cohesive unit: the algorithm selectively extracts the bridging evidence while filtering out semantic noise. 
Consequently, the merged chunk contains only 361 tokens --- significantly more compact than the raw concatenation of the original inputs.
Utilizing this synthesized context, the generator accurately resolves the dependency, determining that Alcala attended \underline{\ding{182} the University of the Philippines}, which initiated military instruction in \underline{\ding{183} 1912}.

\begin{figure}[ht]
    \centering
    \begin{tcolorbox}[
        enhanced,
        colback=white,
        colframe=boxFrame,
        fonttitle=\bfseries,
        title=A Qualitative Analysis Example for \ourname{},
        drop shadow
    ]
    
        \begin{tcolorbox}[
            colback=bgQuery,
            colframe=blue!30!gray,
            title=\small \textbf{Query}, 
            sharp corners,
            boxrule=0.5pt
        ]
            \small
            When did \hlA{military instruction start} at the place where \hlB{Larry Alcala was educated}?
        \end{tcolorbox}
        
        \vspace{0.2cm}
        \centering \large $\swarrow \searrow$
        \vspace{0.1cm}

        \begin{tcbitemize}[
            raster columns=2,
            raster equal height,
            raster force size=false,
            raster column skip=0.2cm
        ]
            \tcbitem[
                colback=bgChunkA,
                colframe=green!30!gray,
                title=\small \textbf{Chunk A (85 tokens)}, 
                fontupper=\footnotesize
            ]   
                \hlB{Larry Alcala earned his Bachelor of Fine Arts in Painting} \hlBridge{at the University of the Philippines (UP)} in 1950. 
                He became a professor at the same university from 1951 to 1981. 
                He also received the Australian Cultural Award accompanied by a travel study grant in 1975. 
                He started his cartooning career in 1946 while still attending school.
               
            \tcbitem[
                colback=bgChunkB,
                colframe=orange!30!gray,
                title=\small \textbf{Chunk B (678 tokens)},
                fontupper=\footnotesize
            ]
                ROTC in the Philippines began \hlA{in 1912 when the Philippine Constabulary commenced with military instruction} \hlBridge{at the University of the Philippines.} 
                The university's Board of Regents then made representations to the Unit. In 1866, by Auditor School, the two-year officers’ courses were established. 
                In 1867, the courses were transformed into a Military Law Academy. 
                \dots 
                
        \end{tcbitemize}

        \vspace{0.1cm}
        \centering \large $\downarrow$ \textit{\footnotesize Merge Algorithm}
        \vspace{0.1cm}

        \begin{tcolorbox}[
            colback=bgResult,
            colframe=gray!50!black,
            title=\small \textbf{Merged Result (361 tokens)},
            boxrule=0.5pt
        ]
            \small
            ROTC in the Philippines began \hlA{in 1912 when the Philippine Constabulary commenced with military instruction} \hlBridge{at the University of the Philippines.}
            The university's Board of Regents then made representations to the Unit. In 1866, by Auditor School, the two-year officers’ courses were established. 
            In 1867, the courses were transformed into a Military Law Academy. 
            \dots
            \hlB{Larry Alcala earned his Bachelor of Fine Arts in Painting at the University of the Philippines (UP) in 1950. }
            He became a professor at the same university from 1951 to 1981. 
            He also received the Australian Cultural Award accompanied by a travel study grant in 1975. He started his cartooning career in 1946 while still attending school.
        \end{tcolorbox}
        \vspace{0.1cm}
        \centering \large $\downarrow$ \textit{\footnotesize Downstream QA Task}
        \vspace{0.1cm}

        \begin{tcolorbox}[
            colback=bgAnswer, 
            colframe=brown!60!black, 
            title=\small \textbf{Answer Derivation},
            boxrule=0.5pt
        ]
            \small
            By traversing the merged context, the model extracts the multi-hop reasoning chain:
            
            \vspace{0.2cm}
            \centering
            \hlB{Larry Alcala} 
            $\xrightarrow{\text{\textit{educated at}}}$ 
            \hlBridge{\mbox{\ding{182}} the University of the Philippines} 
            $\xrightarrow{\text{\textit{military instruction started}}}$ 
            \hlA{\mbox{\ding{183}} 1912}
            
            \vspace{0.2cm}
            \raggedright
            \textbf{Final Answer:} \textbf{1912}
        \end{tcolorbox}
        
    \end{tcolorbox}
\end{figure}